\documentclass[aps,floatfix,amsmath,nofootinbib,amssymb,superscriptaddress]{revtex4}
\usepackage{overpic}
\usepackage{amssymb}
\usepackage{indentfirst}
\usepackage{feynmf}
\usepackage{slashed}
\usepackage{cases}
\usepackage{color}
\usepackage{multirow}
\usepackage{epstopdf}
\usepackage{float}
\usepackage{graphicx,color,bm}
\usepackage{graphics}
\usepackage{amsmath}
\usepackage[colorlinks,
            citecolor=blue,
            anchorcolor=red,
            menucolor=red,
            linkcolor=red,
            filecolor=red,
            runcolor=red,
            urlcolor=blue,
            frenchlinks=red]{hyperref}

\newcommand{\nslash}{\kern 0.2 em n\kern -0.50em /}
\newcommand{\kslash}{\kern 0.2 em k\kern -0.45em /}
\newcommand{\pslash}{\kern 0.2 em p\kern -0.50em /}
\newcommand{\Sslash}{\kern 0.2 em S\kern -0.50em /}
\newcommand{\Pslash}{\kern 0.2 em P\kern -0.50em /}
\newcommand{\Rslash}{\kern 0.2 em R\kern -0.50em /}
\newcommand{\open}{{<\kern -0.3 em{\scriptscriptstyle )}}}

\newcommand{\eps}{\epsilon}
\newcommand{\ii}{{\rm i}}

\newcommand{\de}{d}

\newcommand{\tr}{\text{Tr}}

\begin{document}

\title {Beam single spin asymmetry $A_{LU}^{\sin \phi_R}$ of the dihadron production in SIDIS process}

\author{Yanli Li}
\affiliation{School of Physics, Zhengzhou University, Zhengzhou, Henan 450001, China}
\author{Keyang She}
\affiliation{School of Physics, Zhengzhou University, Zhengzhou, Henan 450001, China}
\author{Hui Li}
\affiliation{School of Physics and Information Engineering, Shanxi Normal University, Taiyuan, Shanxi 030031, China}
\author{Xiaoyu Wang}
\email{xiaoyuwang@zzu.edu.cn (Corresponding Author)}
\affiliation{School of Physics, Zhengzhou University, Zhengzhou, Henan 450001, China}
\author{De-Min Li}
\email{lidm@zzu.edu.cn}
\affiliation{School of Physics, Zhengzhou University, Zhengzhou, Henan 450001, China}
\author{Zhun Lu}
\email{zhunlu@seu.edu.cn}
\affiliation{School of Physics, Southeast University, Nanjing, Jiangsu 211189, China}

\begin{abstract}

We study the longitudinal beam single-spin asymmetry $A_{LU}^{\sin \phi_R}$ of dihadron production in semi-inclusive deep inelastic scattering process with a polarized electron beam scattering off an unpolarized proton target. The asymmetry arises from the convolutions of twist-3 PDF $e(x)$ and twist-2 DiFF $H_1^\sphericalangle$ as well as twist-2 PDF $f_1(x)$ and twist-3 DiFF $\widetilde{G}^{\sphericalangle}$. Using the spectator model calculations for the twist-3 PDF $e(x)$, the DiFFs $D_1 , H_1^\sphericalangle$, and $\widetilde{G}^{\sphericalangle}$, we estimate the BSA $A^{\sin \phi_R}_{LU}$ at the kinematical configuration of CLAS and CLAS12. We find good agreement with $z$- and $M_h$-dependencies data, though $x$- and $Q^2$- dependencies show discrepancies. We include DGLAP evolution of $e(x)$ and provide predictions for COMPASS, EIC, and EicC. The negligible contribution from $f_1(x) \otimes \widetilde{G}^{\sphericalangle}$ supports the Wandzura-Wilczek approximation, while our results highlight the importance of QCD evolution in twist-3 PDF analyses.

\end{abstract}

\maketitle

\section{INTRODUCTION}
\label{Sec.introduce}

Azimuthal asymmetries observed in semi-inclusive deep inelastic scattering (SIDIS) process provide crucial insights into the hadronization mechanism and the partonic structure of hadrons. The SIDIS process is theoretically described by convolutions of parton distribution functions (PDFs) and fragmentation functions (FFs). In recent years, there has been increasing interest in the study of hadron pair production in SIDIS, where the dihadron fragmentation function (DiFF) characterizes the probability of a quark fragmenting into two correlated hadrons. This channel offers direct access to the nucleon structure and allows the investigation of more complicated fragmentation dynamics and correlations~\cite{Metz:2016swz,Wen:2024cfu}.

Higher-twist effects in the SIDIS process encode quark-gluon correlations,  providing an opportunity to study the nucleon beyond the valence structure. Among the twist-3 PDFs, the transverse momentum dependent (TMD) PDF \( e^q(x) \) has attracted considerable interest, as it encodes valuable information related to the nucleon scalar charge, which remains poorly constrained in phenomenological studies~\cite{Boussarie:2023izj,Sharma:2023wha,Jaffe:1991ra,Wakamatsu:2000fd,Wakamatsu:2003uu,Schweitzer:2003uy,Mukherjee:2009uy,Avakian:2010br,Lorce:2014hxa,Pasquini:2018oyz,Bastami:2020rxn}. Furthermore, \( e(x) \) plays a crucial role in understanding the decomposition of the nucleon mass into contributions from quarks and gluons~\cite{Ji:2020baz} via the nucleon sigma terms. Several mass decomposition schemes, derived from the energy-momentum tensor, have been proposed in the literature~\cite{Ji:1994av,Ji:2021mtz,Lorce:2017xzd,Lorce:2021xku}. Particularly, the mass term obtained from the equations of motion for free fields is found to dominate in most quark models~\cite{Lorce:2014hxa}. 
The three contributions to \( e(x) \) can be expressed as:  
\[
e^q(x) = e_{\text{sing.}}^q(x) + e_{qgq}^q(x) + e_{\text{mass}}^q(x).
\]  
At the Electron-Ion Collider (EIC), one of the key goals is to investigate the contribution of the sigma terms to the proton mass and their dependence on mass decomposition schemes~\cite{AbdulKhalek:2021gbh}. 
In addition, the $x^2$ moment of \( e(x) \) is related to the force acting on a transversely polarized quark in an unpolarized target after interacting with the virtual photon~\cite{Burkardt:2008ps}.
Consequently, the access to \( e(x) \) is of significant importance.

The distribution function \( e(x) \) has been calculated in various theoretical frameworks, including the spectator model~\cite{Gamberg:2003pz,Jakob:1997wg}, the chiral quark-soliton model~\cite{Schweitzer:2003uy}, and the bag model~\cite{Avakian:2010br}. As a chiral-odd distribution function, \( e(x) \) has to couple with another chiral-odd function to contribute to an observable in high-energy scattering processes. 
Therefore, $e(x)$ can be extracted from the measurements of twist-3 observables by
HERMES~\cite{HERMES:2006pof}, CLAS\cite{CLAS:2014dmz}, and COMPASS~\cite{Moretti:2019lkw} in single-hadron production SIDIS.
In this case, $e(x)$ is accessible when the transverse momentum of the final hadron is measured and a TMD FF serves as a probe.
Some attempts have been made to extract \( e(x) \) from single-hadron production data~\cite{Efremov:2002ut,CLAS:2003qum}. 
In contrast, measurements of dihadron production in SIDIS allow access to twist-3 observables in the
collinear framework without entangling with other terms in the cross section.
In particular, Ref.~\cite{Courtoy:2022kca} presented the first point-by-point extraction of the twist-3 PDF \( e(x) \) from the dihadron produced SIDIS process.

Accessing \( e(x) \) in dihadron production requires dihadron fragmentation functions (DiFFs) as a probe. DiFFs were first introduced in Ref.~\cite{Konishi:1978yx}, and their evolution equations were subsequently studied in Refs.~\cite{Vendramin:1980wz,Vendramin:1981te}. The study was advanced~\cite{Ceccopieri:2007ip} by incorporating an explicit dependence on the invariant mass \( M_h \) of the hadron pair. The transversely polarized DiFF was then proposed~\cite{Collins:1994ax} to access transversely polarized quarks in the nucleon, leading to the definition of \( H_{1,ot}^{\sphericalangle} \). Ref.~\cite{Bianconi:1999cd} established the basis for all possible DiFFs involving two unpolarized final-state hadrons. In Ref.~\cite{Radici:2001na}, a partial-wave analysis was introduced to clarify the correlation between the two produced hadrons, while the study of subleading-twist DiFFs was further extended by integrating over the transverse momentum~\cite{Bacchetta:2003vn}.

Subsequently, in efforts to access the chiral-odd twist-3 TMD PDF \( e(x) \), attention has been focused on DiFF as a novel and technically viable approach to extract \( e(x) \) from experimental data~\cite{Courtoy:2022kca}. Within this mechanism, the chiral-odd DiFF \( H_1^{\sphericalangle} \)~\cite{Radici:2001na,Bacchetta:2002ux} plays a pivotal role, as it couples with \( e(x) \) in the leading twist in the collinear factorization framework. In parallel, the first extraction of interference fragmentation functions was achieved through \( e^+ e^- \) annihilation process~\cite{Courtoy:2012ry}.
Model calculations for DiFFs have been developed within two different theoretical frameworks: the spectator model~\cite{Bianconi:1999uc,Bacchetta:2006un,Bacchetta:2008wb} and the Nambu–Jona-Lasinio (NJL) quark model~\cite{Matevosyan:2013aka,Matevosyan:2013eia,Matevosyan:2017alv,Matevosyan:2017uls}, providing essential theoretical insights into the dynamical properties of DiFFs.

In Ref.~\cite{Radici:2001na}, the leading-twist differential cross section for polarized SIDIS was presented, featuring various azimuthal modulations involving dihadron fragmentation. This study was later extended to include subleading-twist contributions, as discussed in Refs.~\cite{Bacchetta:2003vn,Diehl:2023nmm}, where the issue of gauge invariance of DiFFs was also addressed. Within this formalism, the structure functions are expressed as convolutions of parton PDFs, DiFFs, and hard scattering coefficients.
More recently, the CLAS and CLAS12 Collaborations~\cite{CLAS:2020igs,Hayward:2021psm} have reported measurements of azimuthal asymmetries in pion-pair production from the scattering of longitudinally polarized electron beams off unpolarized proton targets. In the parton model framework, two sources contribute to the \(\sin\phi_R\) asymmetry~\cite{Bacchetta:2003vn}. The first arises from the coupling of the twist-3 distribution \(e(x)\) with the twist-2 DiFF \(H_1^{\sphericalangle}\), while the second originates from the convolution of the unpolarized distribution function \(f_1(x)\) with the twist-3 DiFF \(\widetilde{G}^{\sphericalangle}\).

In this work, we investigate the \(\sin\phi_R\) asymmetry using spectator model calculations for the twist-3 distribution function \(e(x)\) and the DiFFs \(D_1\), \(H_1^{\sphericalangle}\), and \(\widetilde{G}^{\sphericalangle}\). In addition, we also investigate the impact of the T-odd DiFF $\tilde{G}^{\sphericalangle}$, which encodes the quark-gluon-quark correlation and has not been considered in previous studies. The calculation of \(\widetilde{G}^{\sphericalangle}\) is performed within a spectator model framework following Ref.~\cite{Yang:2019aan}.
To better understand the effects of parton distribution evolution on the asymmetries, we include the Dokshitzer–Gribov–Lipatov–Altarelli–Parisi (DGLAP) evolution of \(e(x)\). Using model inputs for the distribution \(e(x)\)~\cite{Mao:2013waa} and the DiFFs~\cite{Bacchetta:2006un,Yang:2019aan}, we provide numerical estimates of the \(\sin\phi_R\) asymmetry at the kinematical regions of CLAS and CLAS12, and compare our results with the available experimental data. Furthermore, we present predictions for the asymmetry at the kinematics of COMPASS, EIC, and EicC.

The paper is organized as follows:  
In Sec.~\ref{Sec.formalism}, we present the theoretical framework for the \(\sin\phi_R\) azimuthal asymmetry of the dihadron production in SIDIS process, where a longitudinally polarized lepton beam scatters off an unpolarized proton target. In Sec.~\ref{Sec.numerical}, we present numerical estimates of the \(\sin\phi_R\) asymmetries at the kinematics of CLAS and CLAS12, and provide predictions for the asymmetry at the kinematical regions of COMPASS, EicC, and EIC. We summarize our work in Sec.~\ref{Sec.conclusion}.

\section{Formalism of the $A^{\sin\phi_R}_{LU}$ asymmetry of dihadron production in SIDIS}
\label{Sec.formalism}

We present the theoretical framework for the beam single-spin asymmetry $A^{\sin\phi_R}_{LU}$ of dihadron production in semi-inclusive deep inelastic scattering (SIDIS):
\begin{align}
l^\rightarrow(\ell)+p (P)\longrightarrow l(\ell^\prime)+h^+(P_1)+h^-(P_2)+X, \label{eq:process}
\end{align}
where a longitudinally polarized lepton beam scatters off an unpolarized proton target. 
The active quark with momentum $p$ is struck by the virtual photon with momentum $q$, which is given by $\ell- \ell'$. The final-state quark with momentum $k=p+q$ then fragments into two final-state hadrons $h^+$ and $h^-$~(which are a pion pair in this work) with momenta $P_1$ and $P_2$, respectively.
In order to express the differential cross section as well as the physical observables, we adopt the following kinematical variables
\begin{align}
\label{eq:invariants}
&x=\frac{k^+}{P^+},\qquad y=\frac{P\cdot q}{P\cdot l},\qquad z_{i}=\frac{P_{i}^-}{k^-},\nonumber\\
&P_h=P_1+P_2,\qquad z=\frac{P_{h}^-}{k^-}=z_1+z_2, \qquad s=(P+l)^2,\nonumber\\
&R=\frac{P_1-P_2}{2}, \qquad Q^2=-q^2,  \qquad M_h^2={P_h^2}.
\end{align}
Here, $z_i$~($i=1,2$) is the longitudinal momentum fraction of the hadron $h_i$ generated from the fragmented quark. We employ the light-cone coordinates $a^\mu=(a^-, a^+, \bm{a_T})$, where $a^{\pm}=(a^0\pm a^3)/\sqrt{2}$ and $\bm{a}_T$ is the transverse component of the vector. 

\begin{figure}
  \centering
\includegraphics[width=0.6\columnwidth]{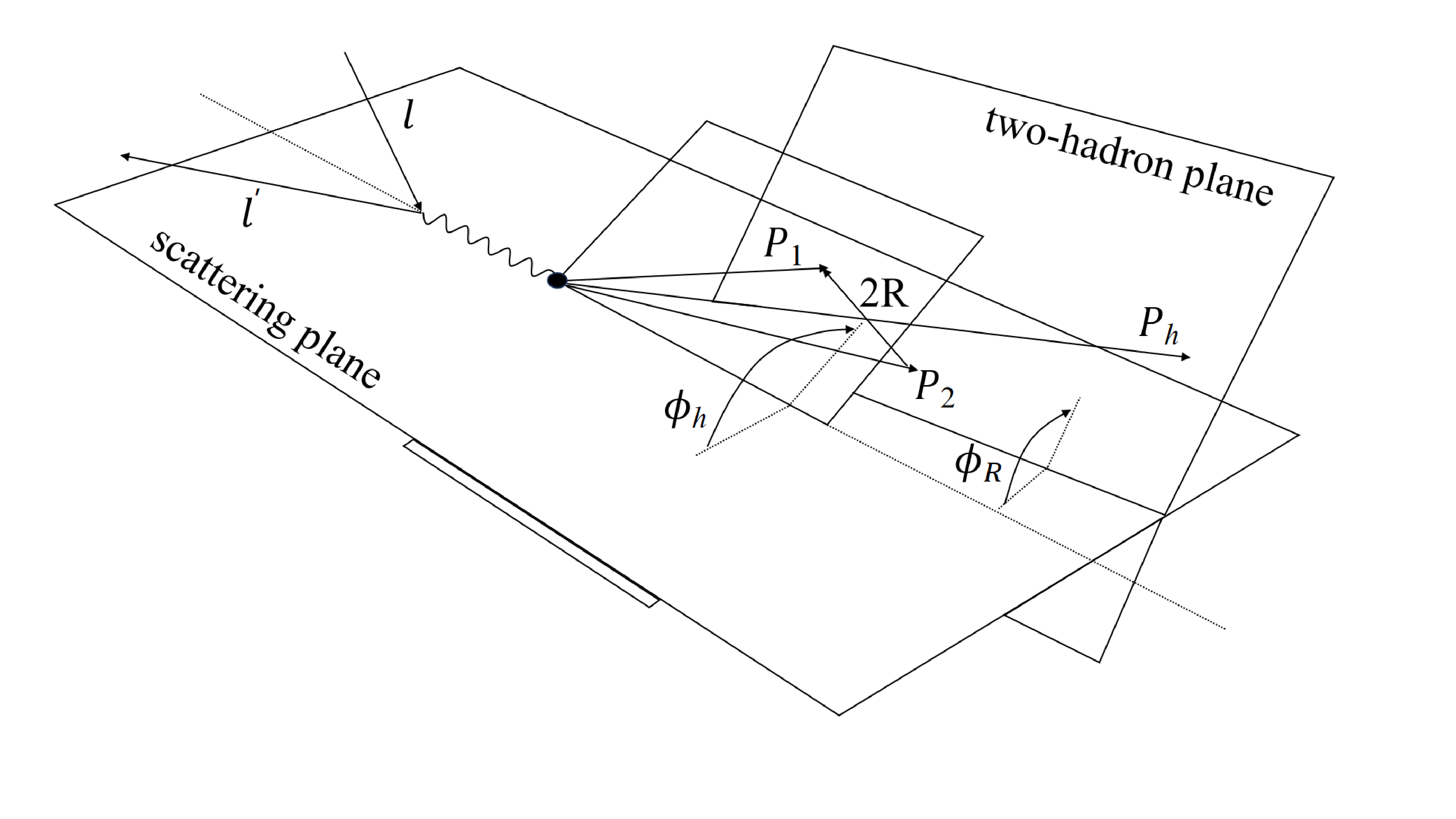}
  \caption{The angle definitions used to measure the beam single spin asymmetry in the dihadron produced SIDIS process.}
  \label{fig:SIDIS-Dihadron}
\end{figure}

With $\theta$ being the polar angle between the direction of $P_1$ in the center-of-mass frame of the hadron pair and the direction of $P_h$ in the lab frame~\cite{Bacchetta:2002ux}, we can express the momenta $P_h^{\mu}$, $k^{\mu}$, and $R^{\mu}$ as follows~\cite{Bacchetta:2006un}
\begin{eqnarray}
P_h^{\mu}&=&\left[P^-_h,\frac{M_h^2}{2P^-_h},\vec{0} \right],\nonumber\\
k^{\mu}&=&\left[\frac{P_h^-}{z},\frac{z(k^2+\vec{k}_T^2)}{2P_h^-},\bm{k}_T \right],\nonumber\\
R^{\mu}&=&\left[\frac{|\vec{R}|P^-_h}{M_h}\cos\theta,-\frac{|\vec{R}|M_h}{2P^-_h}\cos\theta,|\vec{R}|\sin\theta\cos\phi_R,|\vec{R}|\sin\theta\sin\phi_R \right]\nonumber\\
&=&\left[\frac{|\vec{R}|P^-_h}{M_h}\cos\theta,-\frac{|\vec{R}|M_h}{2P^-_h}\cos\theta,R_T^x,R_T^y \right],\nonumber\\
|\vec R| &=& \frac{1}{2} \sqrt{M_h^2 - 2(M_1^2 + M_2^2) + \frac{(M_1^2 - M_2^2)^2}{M_h^2}}\;  \nonumber\\
&=&\frac{M_h}{2}\sqrt{1-\frac{4m^2_\pi}{M^2_h}}.
\end{eqnarray}
Here, $M_1=M_2=m_\pi$ denote the masses of final-state pions,  $\vec{R}_T$ is the component of $R$ that is perpendicular to $P_h$. 

It is convenient to represent the azimuthal angles in the target rest frame, then the angles in the other frames can be obtained by the boost in the direction along $\bm{q}$. The azimuthal angle $\phi_h$ is constructed in the plane transverse to $\bm{q}$ by the angle between the lepton scattering plane and the $(\bm{q},\bm{P_h})$ plane as 
\begin{align}
 \cos \phi_h = \frac{(\hat{q} \times \vec{l})}{|\hat{q} \times \vec{l}|} \cdot \frac{(\hat{q} \times \vec{P_h})}{|\hat{q} \times \vec{P_h}|},
\end{align}
where $\phi_R$ is the angle between the scattering plane and the two-hadron plane (shown in Fig.~\ref{fig:SIDIS-Dihadron}) and has the definition
\begin{align}
 \cos \phi_R = \frac{(\hat{q} \times \vec{l})}{|\hat{q} \times \vec{l}|} \cdot \frac{(\hat{q} \times \vec{R_T})}{|\hat{q} \times \vec{R_T}|}.
\end{align}

In order to perform the partial-wave expansion, the kinematics have been reformulated in the center-of-mass frame of the dihadron system, and several scalar products can be expressed as
\begin{align}
&P_h \cdot R = \frac{M_1^2-M_2^2}{2}=\frac{m_\pi^2-m_\pi^2}{2}=0,  \\
&P_h\cdot k= \frac{M_h^2}{2z}+z\frac{k^2+|\vec{k}_T|^2}{2},\\
&R\cdot k= (\frac{M_h}{2z}-z\frac{k^2+|\vec{k}_T|^2}{2M_h})|\vec{R}|\cos\theta-\bm{k}_T\cdot\bm{R}_T .
\end{align}

 The differential cross section~\cite{Bacchetta:2003vn} for the process in Eq.~(\ref{eq:process}) with a longitudinally polarized lepton beam scattering off an unpolarized proton target can be written as
\begin{align}
\frac{d^6\! \sigma^{}_{LU}}{d\cos \theta\;dM_h^2\;d\phi_R\;dz\;dx\;dy}
=&-\frac{\alpha^2}{Q^2 y}\,2 y \,  \sqrt{1-y} \; \sum_q e_q^2 \,
\frac{M}{Q}\frac{|\bm R|}{M_h}\,\sin{\phi_{R}}\,\nonumber\\
&\times \biggl[x\, e^q(x)\, H_1^{\open, q}\bigl(z,M_h^2,\cos\theta\bigr)
    +\frac{M_h}{Mz}\,f^q_1(x)\,\widetilde{G}^{\open, q}\bigl(z,M_h^2,\cos\theta\bigr)\biggr] \; .
\label{eq:crossLU}
\end{align}
Here, the transverse momentum of the final-state dihadron is integrated out.
Similarly, the differential cross section~\cite{Bacchetta:2003vn} for the unpolarized SIDIS process has the form 
\begin{align}
\frac{d^6\! \sigma^{}_{UU}}{d\cos \theta\;dM_h^2\;d\phi_R\;dz\;dx\;dy} =  \frac{\alpha^2}{ Q^2 y}\,\left(1-y+\frac{y^2}{2}\right)   \sum_q e_q^2 f_1^q(x)\, D_1^q\bigl(z, M_{h}^2, \cos \theta\bigr) . 
\label{eq:crossUU} 
\end{align}
Here, the first and the second subscript in $\sigma_{XY}$ denote the polarization states of the beam and the target, respectively.
In Eq.~(\ref{eq:crossLU}), $e^q(x)$ is the twist-3 distribution function coupled with the twist-2 chiral-odd DiFF $H_{1}^{\sphericalangle,q}\bigl(z,M_h^2,\cos\theta\bigr)$, while the unpolarized PDF $f_1^q(x)$ is coupled with the twist-3 T-odd DiFF $\widetilde{G}^{\sphericalangle,q}\bigl(z,M_h^2,\cos\theta\bigr)$.
In Eq.~(\ref{eq:crossUU}), $D_1^q\bigl(z, M_{hh}^2, \cos \theta\bigr)$ represents the unpolarized DiFF. 

The collinear DiFFs $D_1^q$ and $H_1^{\sphericalangle,q}$ can be extracted from the integrated quark-quark correlator $\Delta (z, R)$ for fragmentation~\cite{Bacchetta:2003vn}
\begin{eqnarray} \label{Eq:6} \nonumber
	\Delta (z, R) \displaystyle &=&z^2 \sum \kern -1.3 em \int_X \;
	\int \frac{d \xi^+}{2\pi} \; e^{\text{i} k\cdot\xi}\;
	\langle 0|\, U^+_{[0,\xi]} \, \psi(\xi) \,|P_h,R; X\rangle
	\langle X; P_h,R|\, \overline{\psi}(0)\,|0\rangle
	\Big|_{\xi^- = \vec{\xi}_T  =0} \; \\
	&=&\frac{1}{16\pi}\left\{ D_1 \slashed{n}_-+ H_{1}^{\sphericalangle}{i\over 2M_h}[R\!\!\!/_T,n\!\!\!/_-]\right\},
\end{eqnarray}
where $U^a_{[b,c]}$ is the gauge-link ensuring the gauge invariance of the operator,  and $n_-$ denotes the negative light-like vector, given by $n_-=[0, 1, \vec{0}_T]$.

The twist-2 fragmentation functions can be derived from the correlation function~\cite{Bacchetta:2002ux}
\begin{equation} 
\Delta^q(z,\cos\theta,M_h^2,\phi_R) = 
\frac{z |\vec R|}{16\,M_h}\int d^2 \vec k_T \; 
       d k^+\,\Delta^q(k;P_h,R) \Big|_{k^- = P_h^-/z}  \; , 
\label{eq:delta1}
\end{equation} 
where $\Delta^q(k,P_h,R)_{ij}$ has the form ~\cite{Boer:2003cm,Bacchetta:2003vn}
\begin{equation} \begin{split} 
\Delta^q(k,P_h,R)_{ij}
         & =\sum \kern -1.3 em \int_X \, \int
        \frac{\de^4\xi}{(2\pi)^{4}}\; e^{+\ii k \cdot \xi}
       \langle 0|
{\cal U}^{n_+}_{(0,-\infty)}
|0\rangle \,.    
\label{e:delta2}
\end{split} 
\end{equation} 

After integrating over the transverse momentum $\vec{k}_T$, we can simplify the Wilson lines ${\cal U}$ to unity by employing a light-cone gauge.  
The two fragmentation functions that survive after $\vec{k}_T$-integration 
are as follows~\cite{Bianconi:1999cd,Bacchetta:2002ux}
\begin{align} 
D_1^q(z,\cos\theta,M_h^2) &= 4\pi\, \tr[\Delta^q(z,\cos\theta,M_h^2,\phi_R)\,
\gamma^-],
\\
\frac{\eps_T^{ij}\,R_{T j}}{M_h}\, H_1^{\open\, q}(z,\cos\theta,M_h^2)
&=4\pi \, \tr[\Delta^q(z,\cos\theta,M_h^2,\phi_R)\,i\,\sigma^{i -}\,\gamma_5].
\end{align} 
These functions can be expanded in terms of the relative partial waves of the pion pair system. By truncating the expansion at the $p$-wave level, we obtain the following results~\cite{Bacchetta:2002ux}
\begin{align}
D_1^{q}(z,\cos{\theta},M_h^2) &\approx D_{1,oo}^{q}(z,M_h^2) + 
D_{1,ol}^{q}(z,M_h^2)\, \cos\theta + D_{1,ll}^{q}(z,M_h^2) \, \frac{1}{4}\,
(3\cos^2\theta -1) \; , 
\label{eq:d1pw}
\\
H_1^{\open\,q}(z,\cos{\theta},M_h^2) &\approx  H_{1,ot}^{\open\,q}(z,M_h^2) + 
H_{1,lt}^{\open\,q}(z,M_h^2) \, \cos\theta \; . 
\label{eq:h1angpw}
\end{align} 
The fragmentation function $D_{1,oo}$ can arise from
both the $s$-wave and $p$-wave contributions, but not from the interference term. $D_{1,ol}$ and $H_{1,ot}^{\open}$ 
originate from the interference of $s$-
and $p$-waves, while $D_{1,ll}$ receives contribution from the polarized $p$-waves. And $H_{1,lt}^{\open}$ originates from the interference of two $p$-waves with
different polarizations.

The twist-3 DiFF $\widetilde{G}^{\sphericalangle}$ emerges from the multiparton correlation during quark fragmentation , which is described by the quark-gluon-quark correlator~\cite{Bacchetta:2003vn,Lu:2015wja}
\begin{eqnarray}
\widetilde{\Delta}_A^{\alpha}(z,k_T,R)&=&\frac{1}{2z}\sum_X\int\frac{d\xi^+d^2\xi_T}{(2\pi)^3}
e^{ik\cdot\xi}\langle0|\int_{\pm\infty^+}^{\xi^+}d\eta^+
{\cal{U}}^{\xi_T}_{(\infty^+,\xi^+)}\cr
&&\times g F_\bot^{-\alpha}{\cal{U}}^{\xi_T}_{(\eta^+,\xi^+)}\psi(\xi)|P_h,R;X\rangle\langle P_h,R;X|\bar{\psi}(0){\cal{U}}^{0_T}_{(0^+,\infty^+)}{\cal{U}}^{\infty^+}_{(0_T,\xi_T)}
|0\rangle\mid_{\eta^+=\xi^+=0,\eta_T=\xi_T}. \label{eq:deltaG}
\end{eqnarray}
Here, $F_\bot^{-\alpha}$ represents the field strength tensor of the gluon. After integrating out $\vec{k}_T$, we obtain the following result
\begin{align}
\widetilde{\Delta}_A^{\alpha}(z,\cos \theta,M_h^2,\phi_R)=\frac{z^2|\vec{R}|}{8M_h}\int d^2\vec{k}_T \widetilde{\Delta}_A^{\alpha}(z,k_T,R).
\label{eq:deltaG1}
\end{align}
Therefore, by taking the trace, we can extract the DiFF $\widetilde{G}^{\sphericalangle}$ from the function $\widetilde{\Delta}_A^{\alpha}(z,k_T,R)$
\begin{align}
\frac{\epsilon_T^{\alpha\beta}R_{T\beta}}{z}\widetilde{G}^{\sphericalangle}(z,\cos \theta,M_h^2)=4\pi\textrm{Tr}[\widetilde{\Delta}_A^{\alpha}(z,\cos \theta,M_h^2,\phi_R)\gamma^{-}\gamma_5].
\end{align}
Similarly, the twist-3 DiFF $\widetilde{G}^{\sphericalangle}$ is obtained by expanding the correlator to the $p$-wave level
\begin{align}
\widetilde{G}^{\sphericalangle}(z,\cos \theta, M_h^2)=\widetilde{G}^{\sphericalangle}_{ot}(z,M_h^2)+\widetilde{G}^{\sphericalangle}_{lt}(z,M_h^2)\cos \theta.
\end{align}
Specifically, $\widetilde{G}^{\sphericalangle}_{ot}$ originates from the interference of $s$- and $p$-waves, and $\widetilde{G}^{\sphericalangle}_{lt}$ comes from the interference of two $p$ waves with different polarizations.

In this study, following the procedure described in Ref.~\cite{Bacchetta:2002ux}, we  neglect the $\cos \theta$-dependent terms in the expansion of DiFFs for two reasons. First, it should be noted that the terms dependent on $\cos \theta$ correspond to higher-order contributions in the partial-wave expansion. Consequently, $\cos \theta$-dependent terms can only be significant when the two hadrons are produced via a spin-1 resonance. Secondly, the $\cos\theta$-dependent terms should vanish with the angular variable $\theta$ being integrated over the interval $[-\pi, \pi]$.
Therefore, we focus on the functions $D^q_{1,oo}$, $e^q(x)$, $H_{1,ot}^{\sphericalangle, q}$, and $\widetilde{G}^{\sphericalangle, q}_{ot}$. In this scenario, the ${\sin\phi_R}$ asymmetry of dihadron production SIDIS process with beam longitudinally polarized and the target unpolarized can be expressed as follows~\cite{Courtoy:2022kca},
\begin{align}
A^{\sin\phi_R}_{LU}(x,z,M_h^2;Q,y) = -\frac{W(y)}{A(y)}\frac{M}{Q}\frac{|\bm{R}|}{M_h}\frac{\sum_{q}e^2_q
\left[xe^q(x)H_{1,ot}^{\sphericalangle,q}(z,M_h^2)+\frac{M_h}{zM} f_1^q(x,Q^2)\widetilde{G}^{\sphericalangle,q}_{ot}(z,M_h^2)\right]}{\sum_{q}e^2_q f^q_1(x,Q^2) D^q_{1,oo}(z,M_h^2)}.
\label{eq:AsinphiR}
\end{align}
Here, the depolarizing factor is given by $W(y)=2 y \sqrt{1-y}$ and $A(y)=1-y+\frac{y^2}{2}$~\cite{Bacchetta:2003vn}.

\section{NUMERICAL CALCULATION}
\label{Sec.numerical}
In this section, we present the numerical estimate of the beam single-spin asymmetry \( A_{LU}^{\sin \phi_R} \), considering the contributions from both the coupling of \( e(x) \) with \( H_1^\sphericalangle \), and the coupling of \( f_1(x) \) with \( \widetilde{G}^{\sphericalangle} \).
In the numerical calculation of the asymmetry, we require the corresponding twist-3 PDF \( e(x) \), DiFF \( \widetilde{G}^{\sphericalangle}_{ot} \), and twist-2 DiFFs \( D_{1,oo} \), \( H_{1,ot}^{\sphericalangle} \) as well as the unpolarized PDF \( f_1(x) \) as necessary inputs. For the unpolarized PDF \( f_1(x) \), we adopt the NLO set of the CT10 parameterization (central PDF set)~\cite{Lai:2010vv}, while for the other three PDF and DiFFs, we use the results of the spectator model, which we will describe in detail.

\subsection{Model calculation of the twist-3 PDF $e(x)$}
In this subsection, we present the calculation of the chiral-odd twist-3 PDF \( e(x) \) within the spectator model, including contributions from both scalar and axial-vector diquarks, following the approach in Refs.~\cite{Mao:2012dk, Mao:2013waa}, to obtain the proton’s \( e(x) \) for both \( u \) and \( d \) quarks. 
Set 2 of Ref.~\cite{Mao:2013waa} has been shown to best reproduce the general trend of the beam single-spin asymmetry in the one-hadron \( \pi^+ \) produced SIDIS process at the kinematical region of CLAS12~\cite{CLAS:2021opg}. 
This model employs a simple propagator \( d_{\mu\nu}(P-k) = -g_{\mu\nu} \) for the axial-vector diquark, and assumes that the axial-vector diquarks (\( uu \) and \( ud \)) are mass-degenerate. The ratio between axial-vector and scalar diquark contributions is fixed by SU(4) spin-flavor symmetry. The contributions from scalar and axial-vector diquarks to \( e(x, \bm{k}_T^2) \), obtained by using the Set 2 spectator model of Ref.~\cite{Mao:2013waa}, are given as follows
\begin{align}
e^{s}(x,\bm k_T^2)&=\frac{1}{16\pi^3} \frac{N_s^2(1-x)^2}{(\bm{k}_T^2+L_s^2)^4}\nonumber\\
& \quad\times \left[(1-x)(xM+m)(M+m)\right.\nonumber\\
& \quad\left.-(1+\frac{m}{M})\bm k_T^2
-(x+\frac{m}{M})M_s^2\right],
 \label{es}
\end{align}

\begin{align}
e^v(x,\bm{k}_T^2)
   &=\frac{N_v^2(1-x)^2}{16\pi^3}
   \frac{1}{(\bm k_T^2+L_v^2)^4}\,\nonumber\\
   & \quad\times \left[(1-x)(M+m)(x M+m)-(2+\frac{m}{M})\bm k_T^2\right.\nonumber\\
   & \quad\left.+(1-x)(m^2+xM^2)-(2x+\frac{m}{M})M_v^2\right].
\label{ev2}
\end{align}
Here, $L_X^2$ has the following form
\begin{align}
L_X^2=(1-x)\Lambda_{X}^2 +x M_{X}^2-x(1-x)M^2,~~ X=s,v,
\end{align}
where $m$, $M_X$ and $M$ represent the quark, diquark, and nucleon masses, respectively. $N_X$ is the coupling constant and $\Lambda_X$ is the cut-off parameter.
With the scalar (\(e^s\)) and axial-vector (\(e^v\)) diquark contributions to \(e(x,\bm{k}_T^2)\), one can construct the distribution functions for the \(u\) and \(d\) valence quarks. However, this procedure involves a degree of freedom in choosing the relation between quark flavors and diquark types.
\begin{table}[h]
\begin{tabular}{|c|c|c|c|c|}
\hline
Diquark &$M_X$ (GeV) & $\Lambda_X$ (GeV) & $c_X^2$ & $N_X^2$ \\
\hline
Scalar $s(ud)$ &0.6 &0.5 &1.5 &6.525 \\
\hline
Axial-vector $a(ud)$ &0.8 &0.5 &0.5 &28.716 \\
\hline
Axial-vector $a'(uu)$ &0.8 &0.5 &1.0 &28.716\\
\hline
\end{tabular}
\caption{Values for the parameters to calculate the DF $e(x,\bm{k}_T^2)$, taken from Ref.~\cite{Bacchetta:2003rz}.}
\label{Tab.1}
\end{table}
As discussed in Refs.~\cite{Bacchetta:2003rz,Jakob:1997wg,Bacchetta:2008af}, a general parametrization for this relation can be expressed as~(where \(f\) denotes a generic TMD PDF)
\begin{equation}
f_u = c_s^2 f_s + c_a^2 f_a, \quad f_d = c_{a'}^2 f_{a'},
\end{equation}
where \(f_a\) and \(f_{a'}\) represent the contributions from the axial-vector isoscalar diquark \(a(ud)\) and the axial-vector isovector diquark \(a'(uu)\), respectively. The coefficients \(c_s\), \(c_a\), and \(c_{a'}\) are model parameters that control the relative weight of each diquark channel. These coefficients are derived from the SU(4) spin-flavor symmetry of the proton wave function, and the masses for different axial diquarks are identical. The values of the parameters involved in this work are presented in Table~\ref{Tab.1}.

\begin{figure*}[h]
  \centering
  \includegraphics[width=0.50\columnwidth]{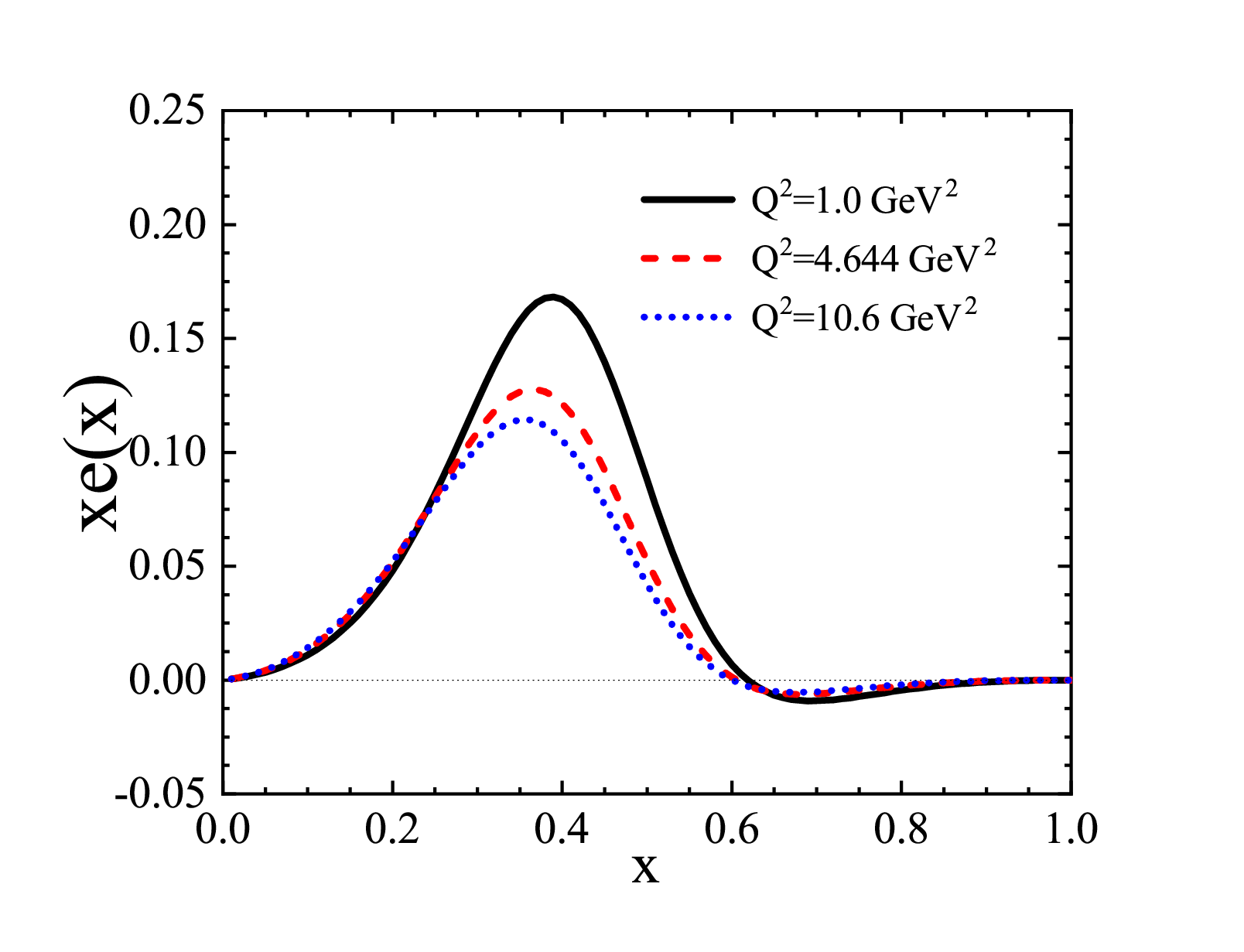} 
  \caption{The \( x \)-dependence of twist-3 PDF $xe(x)$ at different energy scales, $Q^2=1 ~\textrm{GeV}^2$ (black solid line), $Q^2=4.644~\textrm {GeV}^2$ (red dashed line),  and $Q^2=10.6 ~\textrm{GeV}^2$ (blue dotted line).}
  \label{fig:e(x)hoppet}
\end{figure*}

It should be noted that the model results are obtained at a fixed energy scale, whereas the experimental data span a relatively wide range of energy scales. Therefore, it is necessary to take into account the DGLAP evolution effects for the involved PDFs. In this work, we assume that the evolution kernel for the collinear twist-3 PDF \( e(x) \) is the same as that for the unpolarized distribution function \( f_1(x) \), and we perform the DGLAP evolution using the {\sc HOPPET} evolution package~\cite{Salam:2008qg}.
In Fig.~\ref{fig:e(x)hoppet}, we show the \( x \)-dependence of the twist-3 parton distribution function \( x e(x) \) at different energy scales: \( Q^2 = 1~\textrm{GeV}^2 \) (black solid line), \( Q^2 = 4.644~\textrm{GeV}^2 \) (red dashed line), and \( Q^2 = 10.6~\textrm{GeV}^2 \) (blue dotted line).
One can clearly observe the effects of QCD scale evolution on the magnitude of the PDF in its \( x \)-dependence: in the small-\( x \) region~($x<0.25$), the magnitude of \( x e(x) \) increases as the scale evolves from low to high \( Q^2 \), while in the large-\( x \) region~($x>0.25$), it decreases. This behavior suggests that the QCD evolution of the PDF \( e(x) \) can have an impact on the calculation of the \( \sin\phi_R \) azimuthal asymmetry.

\subsection{Model calculation of the DiFFs \( D_{1,oo}^{q} \), \( H_{1,ot}^{\sphericalangle,q} \), and \( \widetilde{G}^{\sphericalangle,q}_{ot} \)}
In this subsection, we present the calculation of the twist-2 DiFFs \( D_{1,oo}^{q} \) and  \( H_{1,ot}^{\sphericalangle,q} \), as well as the twist-3 DiFF \( \widetilde{G}^{\sphericalangle,q}_{ot}\) within the spectator model.

\begin{figure}[t]
\includegraphics[width=6cm]{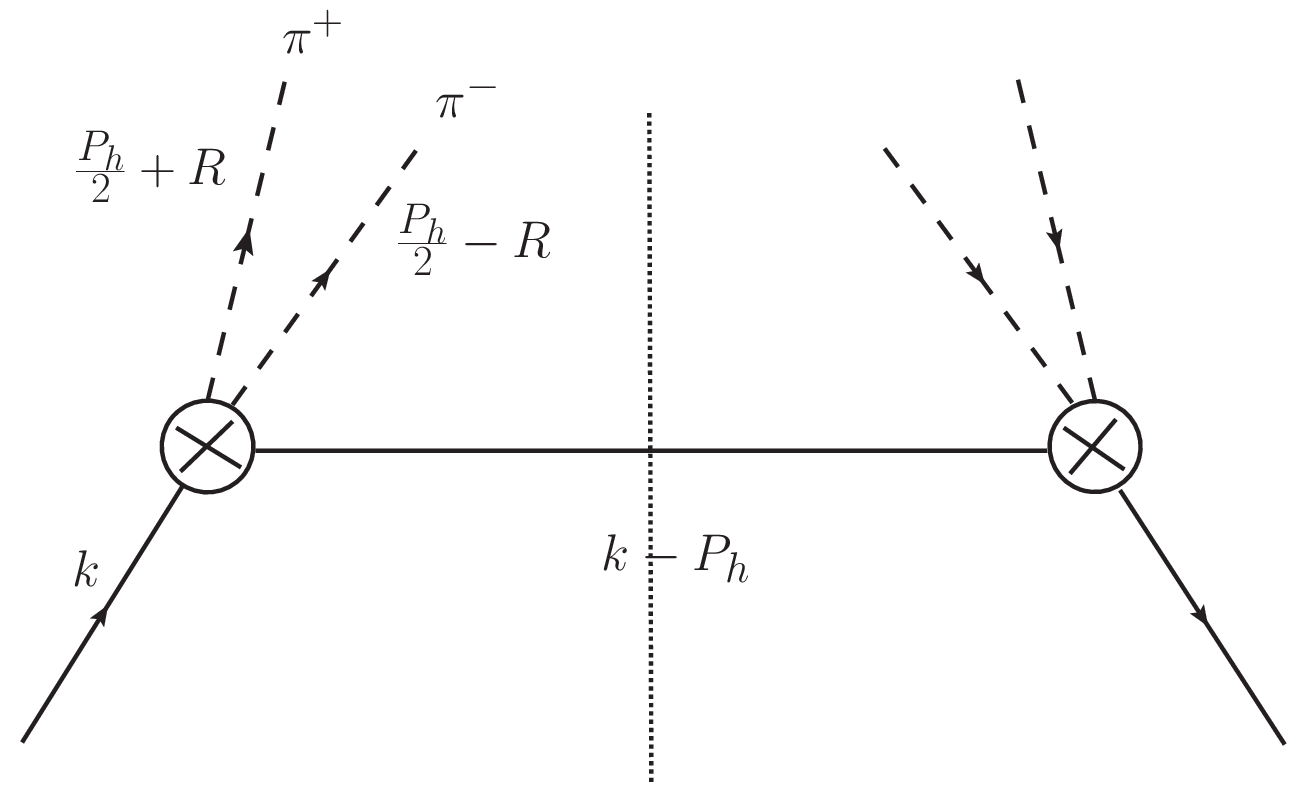}
\caption{
A diagram of the spectator model's correlation function $\Delta$.}
\label{f:defination/spect_fragm.eps} 
\end{figure}

Firstly, we present the calculation of the DiFF \( D_{1,oo}^{q}\) following the spectator model framework in Ref.~\cite{Bacchetta:2006un}, in which the summation over all the possible prominent intermediate states \( X \) in the process $q\to\pi^+\pi^-X$ is replaced by an effective on-shell spectator state.
In principle, different fragmentation channels could correspond to spectators with different masses, and each channel could lead to multiple possible spectator configurations, for simplicity, Ref.~\cite{Bacchetta:2006un} considered a single effective spectator state common to all channels with its mass and momentum as \( M_s \) and \( P_s \).
The correlation function depicted in Fig.~\ref{f:defination/spect_fragm.eps} was modeled as~\cite{Bacchetta:2006un} 
\begin{align} 
\Delta^{q}(k,P_h,R) &=
\frac{1}{(2\pi)^4}\,
 \frac{(\kslash + m)}{k^2 - m^2} \,\Bigl(F^{s \star}\,e^{-\frac{k^2}{\Lambda_s^2}} +
 F^{p \star}\,e^{-\frac{k^2}{\Lambda_p^2}}\,\Rslash\Bigr)
\,
(\kslash - \Pslash_h +M_s)\nonumber\\
& \quad \times 
\Bigl(F^{s}\,e^{-\frac{k^2}{\Lambda_s^2}}+ F^{p}\,
e^{-\frac{k^2}{\Lambda_p^2}}\,\Rslash \Bigr)\,\frac{(\kslash + m)}{k^2 - m^2}\, 
 2\pi\,\delta\bigl((k-P_h)^2 -M_s^2\bigr).
\label{e:deltamod}
\end{align} 
The isospin symmetry implies that the fragmentation correlators for the processes $u \to \pi^+ \pi^- X$, $\bar{d} \to \pi^+ \pi^- X$, $d \to \pi^- \pi^+ X$, and $\bar{u} \to \pi^- \pi^+ X$ are identical. Therefore, the model results for the \( d \) and \( \bar{u} \) quarks can be obtained from those for the \( u \) quark by reversing the sign of \( \vec{R} \), that is, by performing the transformation $\theta \to \pi - \theta$ and $\phi \to \phi + \pi$.
Therefore, the superscript indicating the quark flavor can be dropped, and the fragmentation functions for the process $u \to \pi^+ \pi^- X$ were calculated. Contributions involving the vertex \( F_s \) correspond to the \( s \)-wave, while those with \( F_p \) correspond to the \( p \)-wave component. Exponential form factors are introduced to suppress contributions from large quark virtualities~\cite{Bacchetta:2006un,Gamberg:2003eg}. Alternative choices, such as dipole form factors~\cite{Bianconi:1999uc,Radici:2001na} or sharp cutoffs~\cite{Bacchetta:2002tk}, have also been considered in other studies.

By inserting Eq.~(\ref{e:deltamod}) into Eq.~(\ref{eq:delta1}), one can obtain the following result
\begin{align}  
\Delta(z,\cos\theta,M_h^2,\phi_R)&=\frac{|\vec R|}{128\,\pi^2\,M_h}\,\frac{z^2}{2\,(1-z)\,P_h^-}
\int d |\vec k_T|^2\,\biggl[
|F^s|^2\, e^{-\frac{2\,k^2}{\Lambda_s^2}}\,\frac{(\kslash + m)\,
(\kslash - \Pslash_h +M_s)\,(\kslash + m)}{(k^2 - m^2)^2}\nonumber\\
& \quad+
|F^p|^2\, e^{-\frac{2\,k^2}{\Lambda_p^2}}\,\frac{(\kslash + m)\,\Rslash\;
(\kslash - \Pslash_h +M_s)\,\Rslash\;\,(\kslash + m)}{(k^2 - m^2)^2}\nonumber\\
& \quad +
F^{s \star}\,F^{p}\, e^{-k^2\,\frac{\Lambda_s^2+\Lambda_p^2}{\Lambda_s^2\,\Lambda_p^2}}
\,\frac{(\kslash + m)\,(\kslash - \Pslash_h +M_s)\,\Rslash\;\,(\kslash + m)}{(k^2 - m^2)^2}\nonumber\\
& \quad + F^{s}\,F^{p \star}\, 
e^{-k^2\,\frac{\Lambda_s^2+\Lambda_p^2}{\Lambda_s^2\,\Lambda_p^2}}\,
\frac{(\kslash + m)\,\Rslash\; (\kslash - \Pslash_h +M_s)\,(\kslash + m)}
{(k^2 - m^2)^2} \biggr], \label{e:deltamod2}
\end{align} 
with the value of $k^2$ determined by the on-shell condition of the
spectator~\cite{Bianconi:1999uc}
\begin{equation}
k^2 = \frac{z}{1-z}\,|\vec k_T|^2+\frac{M_s^2}{(1-z)}+\frac{M_h^2}{z} .
\end{equation}
The pure $s$- and $p$-waves contributions are described by the first and second lines of Eq.~(\ref{e:deltamod2}), which are the only ones that can contribute to the functions
$D_{1,oo}$, $D_{1,ll}$ of Eq.~(\ref{eq:d1pw}) and $H_{1,lt}^{\open}$ of
Eq.~(\ref{eq:h1angpw}). The $sp$ interference contributions are described by the third and fourth lines and they contribute to the functions
$D_{1,ol}$ and $H_{1,ot}^{\open}$. The function $L^2$ corresponding to the DiFFs is defined as 
\begin{equation}
L^2(z,M_h^2) = \frac{1-z}{z^2}\,M_h^2 +\frac{1}{z}\,M_s^2 -
\frac{1-z}{z}\,m^2,  
\end{equation} 
which is always positive.
With all ingredients, the model result of the unpolarized fragmentation function $D_{1,oo}$ from the spectator model were presented in Ref.~\cite{Bacchetta:2006un} as

\begin{align}
D_{1,oo}(z,M_h^2) &=  \frac{z\,|\vec R|}{16\, \pi\,M_h}\,|F^s|^2\, e^{-\frac{2\,m^2}{\Lambda_s^2}}\,
\Biggl[\biggl(1+2\,\frac{M_h^2-(m+M_s)^2}{z\,\Lambda_s^2}\biggr)
\,\Gamma\biggl(0,\frac{2\,z\,L^2}{(1-z)\,\Lambda_s^2}\biggr) 
\nonumber\\
& \quad - \frac{1-z}{z^2}\,\frac{M_h^2-(m+M_s)^2}{L^2}\, e^{-\frac{2\,z\,L^2}{(1-z)\,\Lambda_s^2}}\Biggr] \nonumber\\
& \quad + \frac{z\,|\vec R|}{16\, \pi\,M_h}\,|F^p|^2\, e^{-\frac{2\,m^2}{\Lambda_p^2}}\,
\frac{|\vec R|^2}{3\,M_h^2}
\,\Biggl[\biggl(2\,M_h^2+\frac{2-z}{z}\,(m^2-M_s^2)\nonumber
\\ 
& \quad + 2\,\frac{(M_h^2 - (m-M_s)^2)\,(2\,M_h^2 + (m+M_s)^2)}{z\,\Lambda_p^2}
\biggr)\,\Gamma\biggl(0,\frac{2\,z\,L^2}{(1-z)\,\Lambda_p^2}\biggr)\nonumber\\ 
& \quad +\frac{1-z}{2\,z^2\,L^2}\,\biggl(\Bigl(M_s^2 + \frac{1-z}{z}\,(M_h^2-z\, m^2)\Bigr)\Lambda_p^2\nonumber
\\ 
& \quad-2\,(M_h^2 - (m-M_s)^2)\,(2\,M_h^2 + (m+M_s)^2)\biggr)\, e^{-\frac{2\,z\,L^2}{(1-z)\,\Lambda_p^2}}
\Biggr]. 
\end{align}  
The first term of the fragmentation function can be identified with the pure
$s$-wave contribution, also referred to as $D_{1,oo}^{s}/4$ in
Ref.~\cite{Bacchetta:2002ux}, while the second term corresponds to the pure $p$-wave contribution, also known as $3\, D_{1,oo}^{p}/4$.

The vertices \( F_s \) and \( F_p \) play crucial roles in reproducing the dependence of the invariant mass and generating the imaginary parts necessary for the emergence of the T-odd fragmentation function $H_{1,ot}^{\open}$. The $s$-wave component was assumed to be free of resonances, thus \( F_s \) can be treated as a real function at the tree level. The interference fragmentation function $H_{1,ot}^{\open}$ was obtained as~\cite{Bacchetta:2006un}
\begin{equation} \begin{split} 
H_{1,ot}^{\open}(z,M_h^2) &= -\frac{z\,|\vec R|}{16\, \pi\,M_h}\,2\,{\rm
  Im}(F^{s \star} \, F^p)\, 
e^{-\frac{2\,m^2}{\Lambda_{sp}^2}}\,
\,\frac{M_h}{z^2}\, 
\Biggl[
\frac{1-z}{z}\,\frac{M_h^2-z^2\,m^2}{L^2}\,
e^{-\frac{2\,z\,L^2}{(1-z)\,\Lambda_{sp}^2}}
\\ & \quad 
-\biggl(z +2\,\frac{M_h^2-z^2\,m^2}{\Lambda_{sp}^2}\biggr)\,
        \Gamma\biggl(0,\frac{2\,z\,L^2}{(1-z)\,\Lambda_{sp}^2}\biggr) 
\Biggr].
\label{e:h1angle}
\end{split} \end{equation}  
where the function $\Gamma(0,z) \equiv
\int_z^\infty e^{-t}/t\, dt$ is defined in Ref.~\cite{Yang:2017cwi}, which represents the incomplete $\Gamma$ function typically appearing in model calculations involving exponential form-factors~\cite{Gamberg:2003eg}. 
Eq.(\ref{e:h1angle}) demonstrates that when $F^s$ is assumed to be real, $H_{1,ot}^{\open}$ is proportional to the imaginary component of $F^p$, that is, ${\rm Im}(F^p)$.

\begin{figure}[h]
  \centering
  \includegraphics[width=0.4\columnwidth]{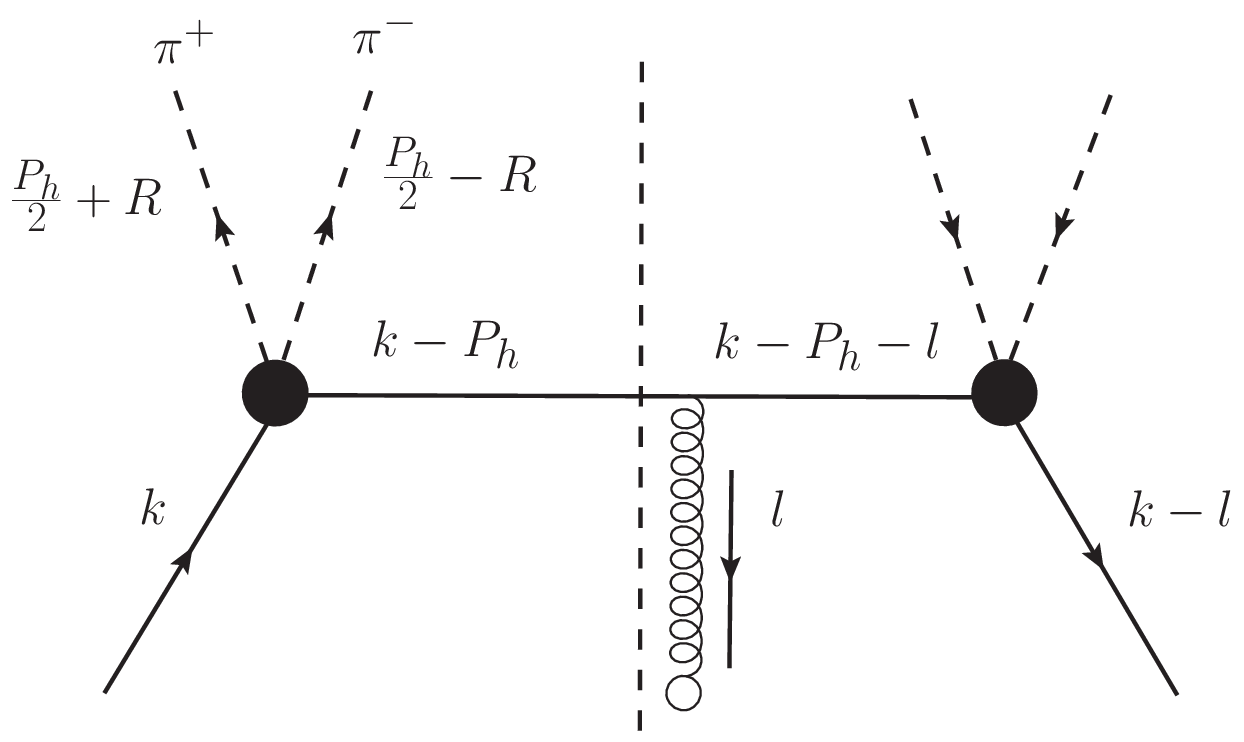}
  \caption{A diagram of the correlation function $\widetilde{\Delta}_A^{\alpha}$ in the spectator model.}
  \label{defination/Fig2.eps}
\end{figure}

For the twist-3 dihadron fragmentation function \( \widetilde{G}^{\sphericalangle}_{ot} \), we adopt the spectator model from Ref.~\cite{Yang:2019aan}. Unlike the twist-2 T-odd DiFF \( H^{\sphericalangle}_{1,ot} \), the DiFF \( \widetilde{G}^{\sphericalangle}_{ot} \) originates from the quark-gluon-quark correlation at twist-3 level, where the degree of freedom of the gluon appears explicitly, as shown in the operator definition in Eq.~(\ref{eq:deltaG}). 
The diagrammatic representation of the $qgq$ correlation function is presented in Fig.~\ref{defination/Fig2.eps}. The left-hand side of Fig.~\ref{defination/Fig2.eps} corresponds to the quark-hadron vertex \( \langle P_h; X| \bar{\psi}(0)|0 \rangle \), while the right-hand side involves the vertex that includes gluon rescattering, \( \langle 0|igF^{- \alpha}_{\perp}(\eta^+)\psi(\xi^+)|P_h; X \rangle \). Here, the Feynman gauge is used, in which case the transverse gauge links \( U^{\xi_T} \) and \( U^{0_T} \) can be neglected~\cite{Belitsky:2002sm, Boer:2003cm}. Within the spectator model framework, the \(s\)-wave and \(p\)-wave contributions to the $qgq$ correlator for dihadron fragmentation can be expressed as follows
\begin{align}
\widetilde{\Delta}_A^{\alpha}(k,P_h,R)&=i\frac{C_F\alpha_s}{2(2\pi)^2(1-z)P_h^-}\frac{1}{k^2-m^2}\int \frac{d^4l}{(2\pi)^4}(l^-g_T^{\alpha\mu}-l^\alpha_T g^{-\mu})(k\!\!\!/ -l\!\!\!/ +m)\nonumber\\
&\quad \times\frac{(F^{s\star} e^{-\frac{k^2}{\Lambda_s^2}}+F^{p\star} e^{-\frac{k^2}{\Lambda_p^2}} R\!\!\!/)(k\!\!\!/ -P\!\!\!/_h-l\!\!\!/+m_s)\gamma_{\mu}(k\!\!\!/ -P\!\!\!/_h+m_s)(F^{s} e^{-\frac{k^2}{\Lambda_s^2}}+F^{p} e^{-\frac{k^2}{\Lambda_p^2}} R\!\!\!/)(k\!\!\!/+m)}{(-l^-\pm i\epsilon)((k-l)^2-m^2-i\epsilon)((k-P_h-l)^2-m_s^2-i\epsilon)(l^2-i\epsilon)}, \label{Delta1}
\end{align}
where $m$ and $m_s$ represent the masses of the fragmenting quark and the spectator, respectively, the factor $(l^-g_T^{\alpha \mu} -l_T^\alpha g^{-\mu})$ is derived from the Feynman rule associated with the gluon field strength tensor. 
With the correlator, the final result for $\widetilde{G}^{\sphericalangle}_{ot}(z,M_h^2)$ is given as
\begin{align}
\widetilde{G}^{\sphericalangle}_{ot}(z,M_h^2)& =\frac{\alpha_s C_F z^2 |\vec{R}|}{8 (2\pi)^4 (1-z)M_h}\frac{1}{k^2-m^2}\int d|\vec{k}_T|^2 e^{-\frac{2 k^2}{\Lambda_{sp}^2}}
\bigg{\{}  \textrm{Im}(F^{s*}F^p) \,C  \nonumber\\
&\quad+ \textrm{Re}(F^{s*}F^p)(k^2-m^2)m_s \big{[}(A + z B) - I_2 \big{]}  \bigg{\}}\,. \label{Gtilde}
\end{align}

The coefficient $C$ has the following form
\begin{align}
&C=m\int_0^1 dx \int_0^{1-x} dy \frac{-2\left[(x+y)k\cdot P_h-y M_h^2\right] +(k^2-m^2) }{x(1-x)k^2+2k\cdot(k-P_h)xy+x m^2 +y^2 m_s^2}\,,
\end{align}
which is proportional to the fragmenting quark mass $m$. The coefficients $A$ and $B$ can be derived from the decomposition of the integral~\cite{Lu:2015wja,Yang:2016mxl}
\begin{align}
&\int d^4l { l^\mu\, \delta(l^2)\, \delta((k-l)^2-m^2)\over (k-P_h-l)^2-m_s^2}=A\, k^\mu + B\, P_h^\mu\,,
\end{align}
with the following expressions
\begin{align}
A&={I_{1}\over \lambda(M_h,m_s)} \left(2k^2 \left(k^2 - m_s^2 - M_h^2\right) {I_{2}\over \pi}+\left(k^2+M_h^2 - m_s^2\right)\right), \\
B&=-{2k^2 \over \lambda(M_h,m_s) } I_{1}\left (1+{k^2+m_s^2-M_h^2 \over \pi} I_{2}\right)\,.
\end{align}
The functions $I_{i}$ are defined as follows~\cite{Amrath:2005gv}
\begin{align}
I_{1} &=\int d^4l \delta(l^2) \delta((k-l)^2-m^2) ={\pi\over 2k^2}\left(k^2-m^2\right)\,, \\
I_{2} &= \int d^4l { \delta(l^2) \delta((k-l)^2-m^2)\over (k-P_h-l)^2-m_s^2}
={\pi\over 2\sqrt{\lambda(M_h,m_s)} }  \ln\left(1-{2\sqrt{ \lambda(M_h,m_s)}\over k^2-M_h^2+m_s^2 + \sqrt{ \lambda(M_h,m_s)}}\right)\,.
\end{align}
Here, $\lambda(M_h,m_s)=(k^2-(M_h+m_s)^2)(k^2-(M_h-m_s)^2)$.

There are several parameters during the model calculation of the DiFFs, which we will present below.
The $s$-wave vertex is assumed to be real in the spectator model we adopted, while the $p$-wave amplitude incorporated contributions from the $\rho$ and $\omega$ mesons, which have the expressions as 
\begin{align}
F^s &= f_{s},
\\
\begin{split} 
F^p &= f_{\rho}\,\frac{(M_h^2 - M_{\rho}^2)- i\, \Gamma_{\rho}\,M_{\rho}}
        {(M_h^2 - M_{\rho}^2)^2 + \Gamma_{\rho}^2\,M_{\rho}^2}
+ f_{\omega}\,\frac{(M_h^2 - M_{\omega}^2)- i\, \Gamma_{\omega}\,M_{\omega}}
        {(M_h^2 - M_{\omega}^2)^2 + \Gamma_{\omega}^2\, M_{\omega}^2}
\\ & \quad
-
i\,f'_{\omega}\,\frac{\sqrt{\lambda\bigl(M_{\omega}^2,M_h^2,m_{\pi}^2\bigr)}\,
\Theta(M_{\omega}-m_{\pi}-M_h)}
        {4\,\pi\,\Gamma_{\omega}\,M_{\omega}^2\,
        \sqrt[4]{4\,M_{\omega}^2\,m_{\pi}^2
        +\lambda\bigl(M_{\omega}^2,M_h^2,m_{\pi}^2\bigr)}},
\label{e:Fp}
\end{split}
\end{align} 
where the couplings $f_{s}$, $f_{\rho}$, $f_{\omega}$ and $f'_{\omega}$, are model parameters. $\Theta$ denotes the unit step function, and $\lambda\bigl(M_{\omega}^2,M_h^2,m_{\pi}^2\bigr) = [M_{\omega}^2 -
(M_h+m_{\pi})^2][M_{\omega}^2 - (M_h-m_{\pi})^2]$. The masses and widths of the $\rho$ and $\omega$ resonances are taken from the Particle Data Group (PDG)~\cite{Singh:2005sbx}: $M_{\rho} = 0.776$
GeV, $\Gamma_{\rho} = 0.150$ GeV, $M_{\omega} = 0.783$
GeV, $m_{\pi} = 0.140$
GeV, $\Gamma_{\omega}  = 0.008$ GeV. 
The $z$-dependent $\Lambda$-cutoff parameters are denoted as
\begin{equation}
\Lambda_{s,p}= \alpha_{s,p}\, z^{\beta_{s,p}}\,(1-z)^{\gamma_{s,p}}, \,
\frac{2}{\Lambda_{sp}^2}=
\frac{1}{\Lambda_{s}^2}+\frac{1}{\Lambda_{p}^2}.
\end{equation} 
The values for the parameters $m$, $m_s$, $\alpha_{s,p}$, $\beta_{s,p}$, and $\gamma_{s,p}$ were adopted from Ref.~\cite{Bacchetta:2006un}, where the model parameters were tuned to the output of the PYTHIA event generator for dihadron production in SIDIS, which can reproduce the unpolarized data well

\begin{align}
\alpha_s& = 2.60~ \mathrm{GeV}^2\,,\qquad  \beta_s = -0.751\,,\qquad \gamma_s = -0.193 \nonumber\,,\\
\alpha_p& = 7.07~ \mathrm{GeV}^2\,,\qquad  \beta_p = -0.038\,,\qquad \gamma_p = -0.085 \nonumber\,,\\
f_s& = 1197~ \mathrm{GeV}^{-1}\,,\qquad  f_\rho = 93.5\,,\qquad f_\omega = 0.63 \nonumber\,,\\
f'_\omega& = 75.2 \,,\qquad  M_s = 2.97 M_h\,,\qquad m = 0.0~\mathrm{GeV} (\textrm{fixed})\,.
\end{align}
In particular, the quark mass $m$ is fixed at $0$ GeV. Consequently, in this model, the twist-3 DiFF $\widetilde{G}^{\sphericalangle}_{ot}$ is derived from the contribution of the $\textrm{Re}(F^{s*}F^p)$ term in Eq.~(\ref{Gtilde}) numerically, as the $\textrm{Im}(F^{s*}F^p)$ term is proportional to the quark mass $m$. As for the strong coupling, we choose $\alpha_s\approx0.3$.

In our analysis, we make the following assumptions
(valid only for $\pi^+ \pi^-$ pairs) based on the isospin symmetry and charge conjugation~\cite{Bacchetta:2006un,Courtoy:2014ixa,Courtoy:2022kca} for twist-2 DiFFs $D_1^q$, $H_1^{\sphericalangle,q}$, and twist-3 DiFF $\widetilde{G}^{\sphericalangle,q}_{ot}$
\begin{align}
D_1^u&=D_1^d=D_1^{\bar{u}}=D_1^{\bar{d}},\\
H_1^{\sphericalangle,u}&=-H_1^{\sphericalangle,d}=-H_1^{\sphericalangle,\bar{u}}=H_1^{\sphericalangle,\bar{d}},\\
\widetilde{G}^{\sphericalangle,u}_{ot}&=-\widetilde{G}^{\sphericalangle,d}_{ot}=-\widetilde{G}^{\sphericalangle,\bar{u}}_{ot}=\widetilde{G}^{\sphericalangle,\bar{d}}_{ot}.
\end{align}

\subsection{Numerical estimate of the BSA $A_{LU}^{\sin \phi_R}$}

In the following, based on the formalism established above, we numerically estimate the $\sin\phi_R$ azimuthal asymmetry of the dihadron production in SIDIS process, where a longitudinally polarized electron beam scatters off an unpolarized proton target. The calculation is performed at the kinematical regions of CLAS, CLAS12, COMPASS, as well as the future facilities EIC and EicC, by taking into account both the $e\, H_{1,ot}^{\sphericalangle}$ term and the $f_1\, \widetilde{G}^{\sphericalangle}_{ot}$ term. Using Eq.~(\ref{eq:AsinphiR}), we derive the expressions for the $x$-, $z$-, $M_h$-, $y$- and $Q^2$-dependent $\sin\phi_R$ asymmetries as follows

\begin{align}
\label{Ax}
A^{\sin\phi_R}_{LU}(x)&=-\frac{\int dy \int dz \int  dM_h\,2M_h \,W(y)\,\frac{|\vec{R}|}{Q}
\frac{|M|}{M_h}\,[(4e^u(x)-e^d(x))xH_{1,ot}^{\sphericalangle}(z,M_h^2)+\frac{M_h}{zM} (4 f_1^u(x)-f_1^d(x))\widetilde{G}^{\sphericalangle}_{ot}(z,M_h^2)]
}{\int dy \int dz \int dM_h\,2M_h  \,A(y)\,[(4 f_1^u(x)+f_1^d(x))D_{1,oo}(z,M_h^2)]},\\
\label{Az}
A^{\sin\phi_R}_{LU}(z)&=-\frac{\int dy \int dx\int dM_h\,2M_h \,W(y)\,\frac{|\vec{R}|}{Q}
\frac{|M|}{M_h}\,[(4e^u(x)-e^d(x))xH_{1,ot}^{\sphericalangle}(z,M_h^2)+\frac{M_h}{zM} (4 f_1^u(x)-f_1^d(x))\widetilde{G}^{\sphericalangle}_{ot}(z,M_h^2)]
}{\int dy \int dx \int dM_h\,2M_h \, A(y)\,[(4 f_1^u(x)+f_1^d(x))D_{1,oo}(z,M_h^2)]},
\end{align}
\begin{align}
\label{Amh}
A^{\sin\phi_R}_{LU}(M_h)&=-\frac{\int dy \int dx\int dz \,
W(y)\, \frac{|\vec{R}|}{Q}
\frac{|M|}{M_h}\,[(4e^u(x)-e^d(x))xH_{1,ot}^{\sphericalangle}(z,M_h^2)+\frac{M_h}{zM} (4 f_1^u(x)-f_1^d(x))\widetilde{G}^{\sphericalangle}_{ot}(z,M_h^2)]
}{\int dy \int dx\int dz \,A(y)\,[(4 f_1^u(x)+f_1^d(x))D_{1,oo}(z,M_h^2)]},\\
\label{Ay}
A^{\sin\phi_R}_{LU}(y)&=-\frac{\int dx\int dz \,\int dM_h \,2M_h \,W(y)\, \frac{|\vec{R}|}{Q}
[\frac{|M|}{M_h}\,[(4e^u(x)-e^d(x))xH_{1,ot}^{\sphericalangle}(z,M_h^2)+\frac{M_h}{zM}(4 f_1^u(x)-f_1^d(x))\widetilde{G}^{\sphericalangle}_{ot}(z,M_h^2)]
}{\int dx\int dz \, \int dM_h \,2M_h  \,A(y)\,[(4 f_1^u(x)+f_1^d(x))D_{1,oo}(z,M_h^2)]}.
\end{align}
\begin{align}
\label{AQ^2}
A^{\sin\phi_R}_{LU}(Q^2)&=-\frac{\int dx\int dz \,\int dM_h \,2M_h  \frac{1}{x} \,W(y)\, \frac{|\vec{R}|}{Q}
\frac{|M|}{M_h}\,[(4e^u(x)-e^d(x))xH_{1,ot}^{\sphericalangle}(z,M_h^2)+\frac{M_h}{zM}(4 f_1^u(x)-f_1^d(x))\widetilde{G}^{\sphericalangle}_{ot}(z,M_h^2)]
}{\int dx\int dz \, \int dM_h \,2M_h \frac{1}{x}\,A(y)\,[(4f_1^u(x)+f_1^d(x))D_{1,oo}(z,M_h^2)]}.
\end{align}
\begin{figure*}
  \centering
  \includegraphics[width=0.4\columnwidth]{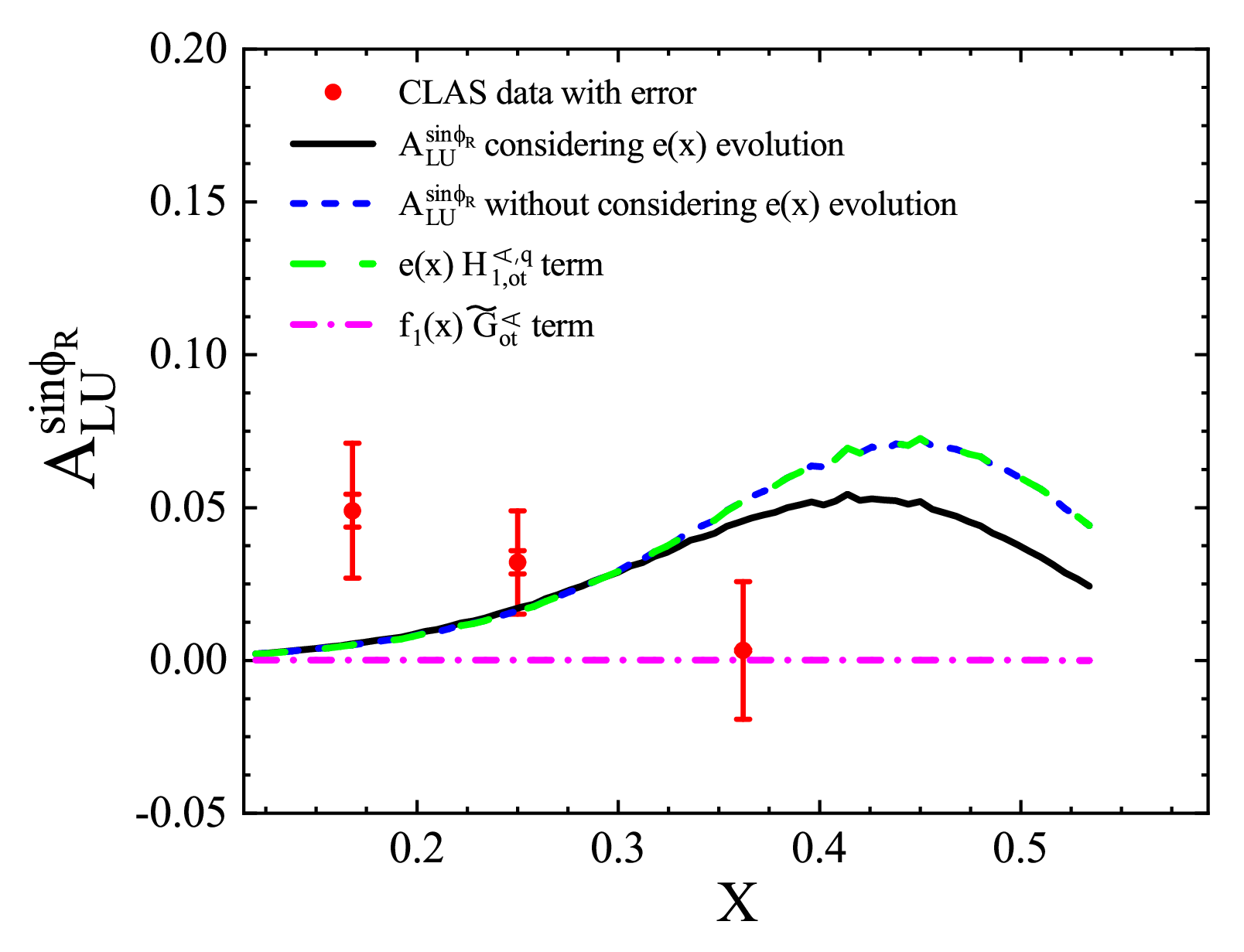}
  \includegraphics[width=0.4\columnwidth]{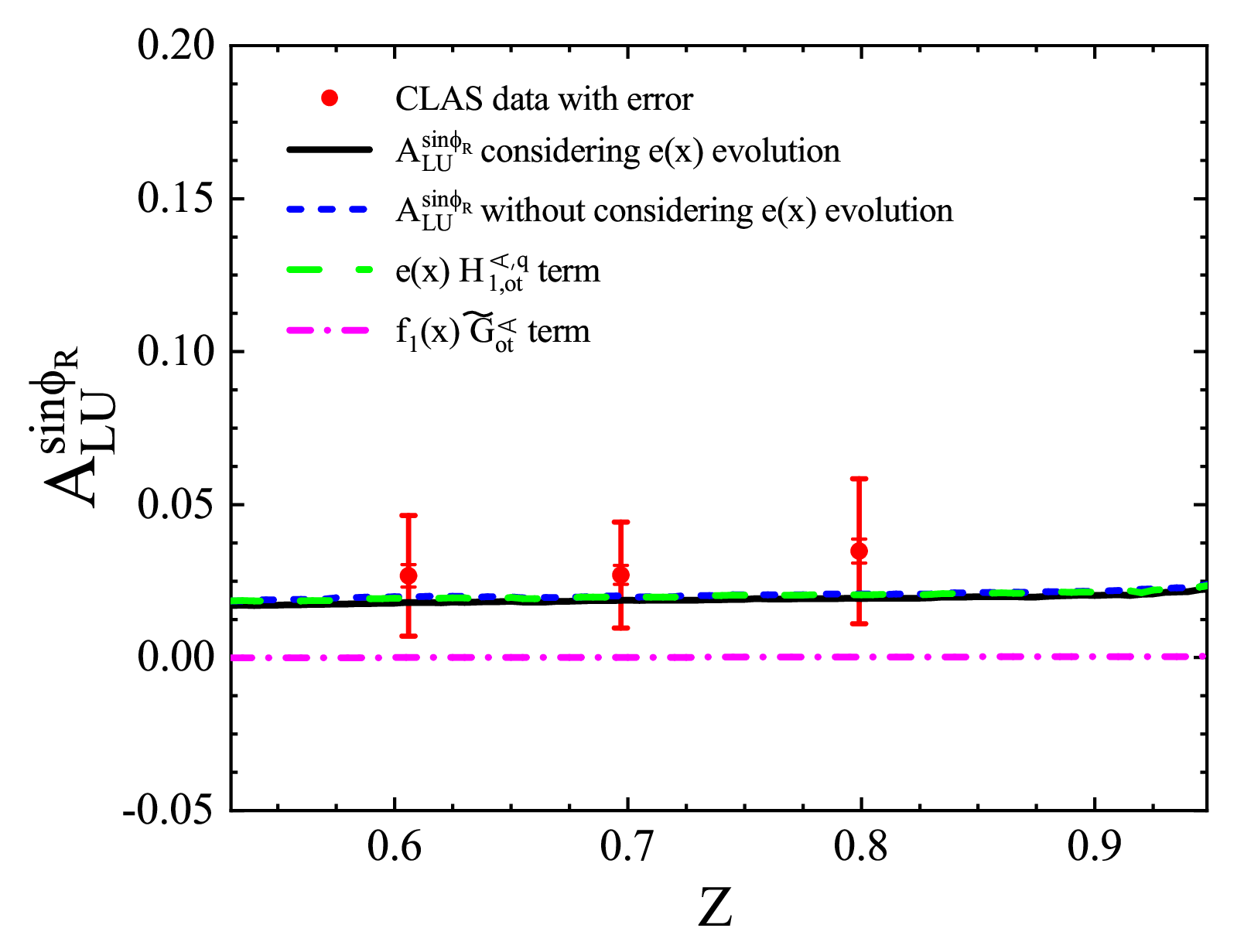}
  \includegraphics[width=0.4\columnwidth]{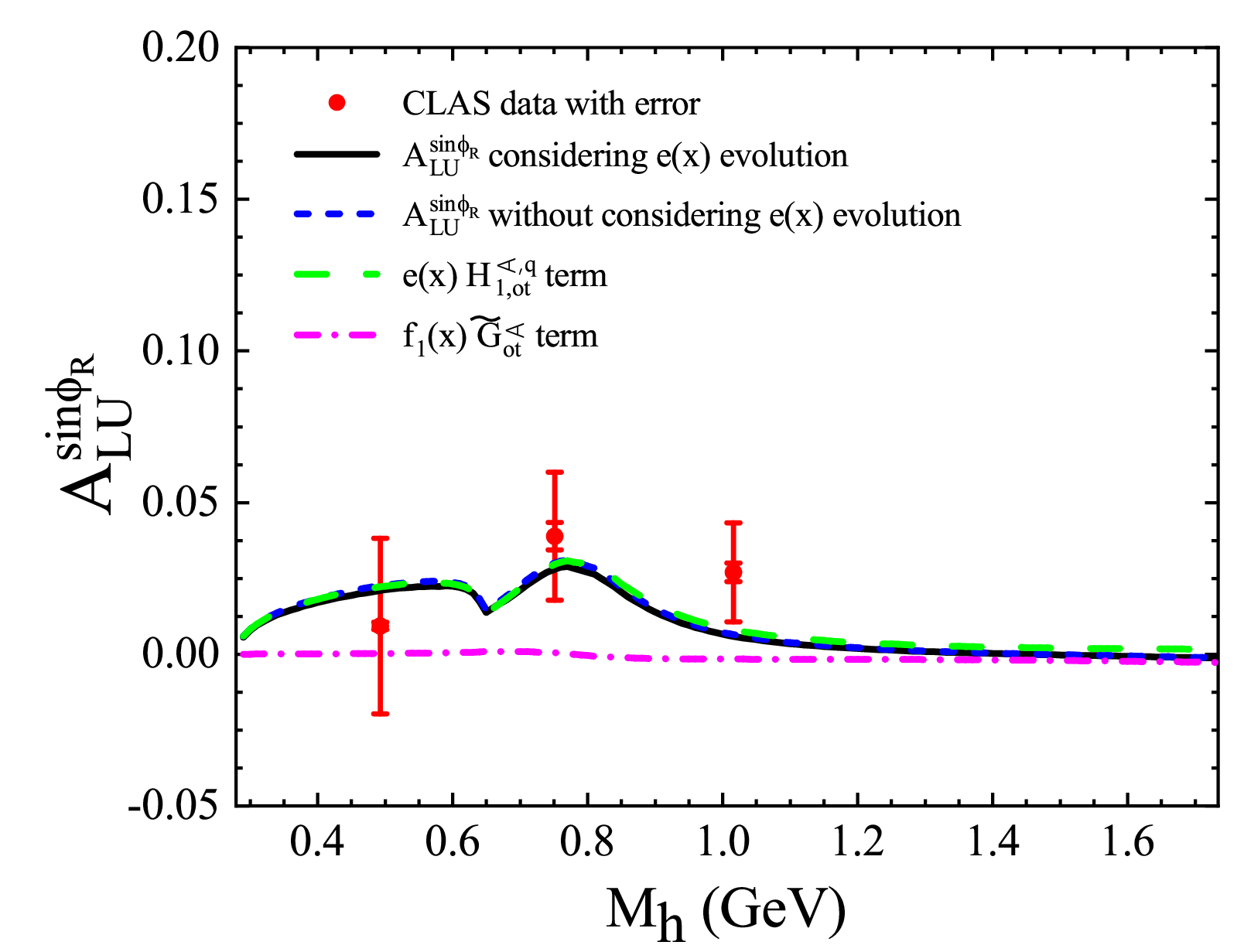}
  \includegraphics[width=0.4\columnwidth]{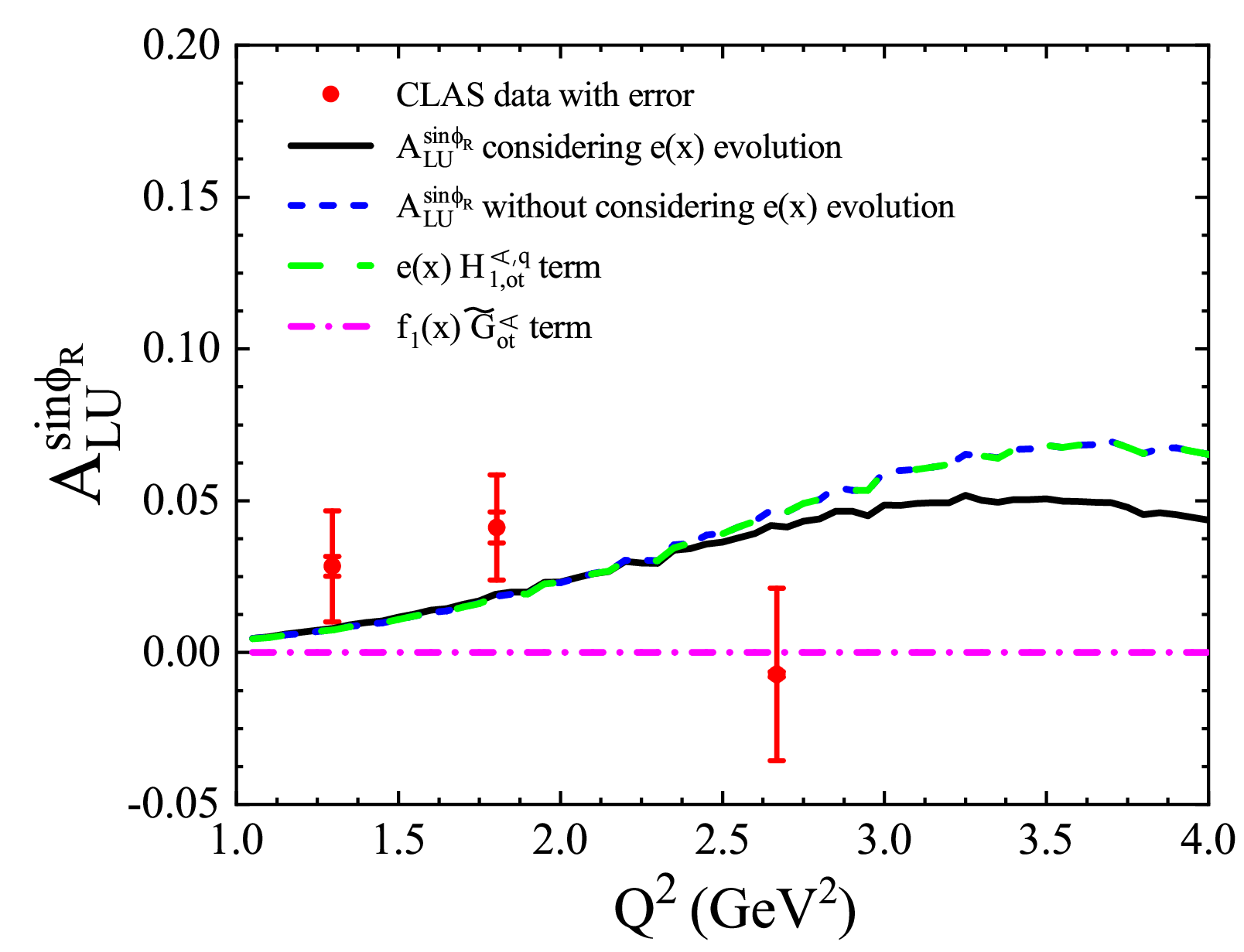}
  \caption{
The $\sin\phi_R$ azimuthal asymmetry of the dihadron production as functions of $x$ (upper left panel), $z$ (upper right panel), $M_h$ (lower left panel), and $Q^2$ (lower right panel) at  CLAS. 
The full circles with error bars represent the experimental data from the CLAS Collaboration~\cite{CLAS:2020igs} for comparison.}
  \label{fig:asyCLAS}
\end{figure*}

\begin{figure*}[h]
  \centering
  \includegraphics[width=0.4\columnwidth]{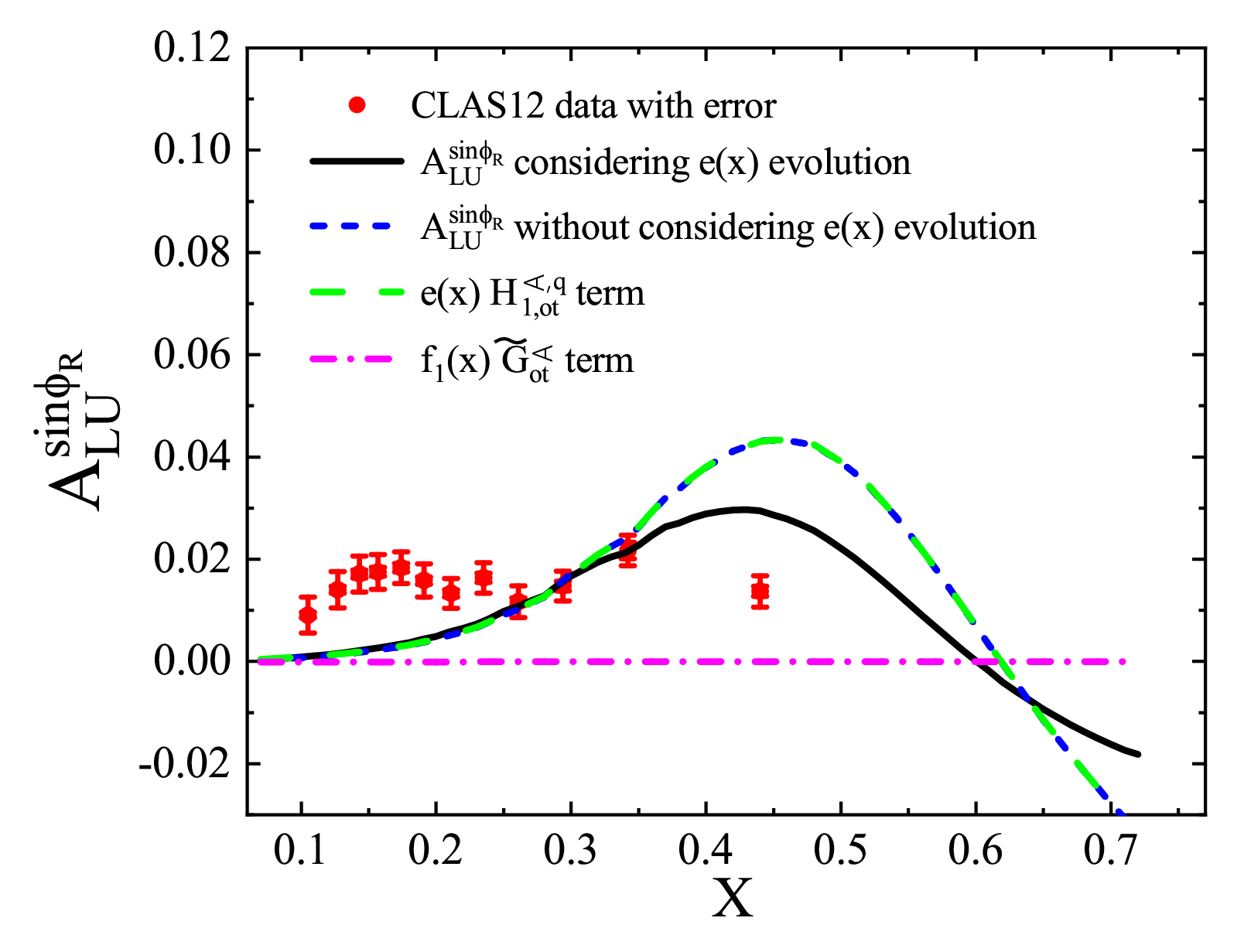}
  \includegraphics[width=0.4\columnwidth]{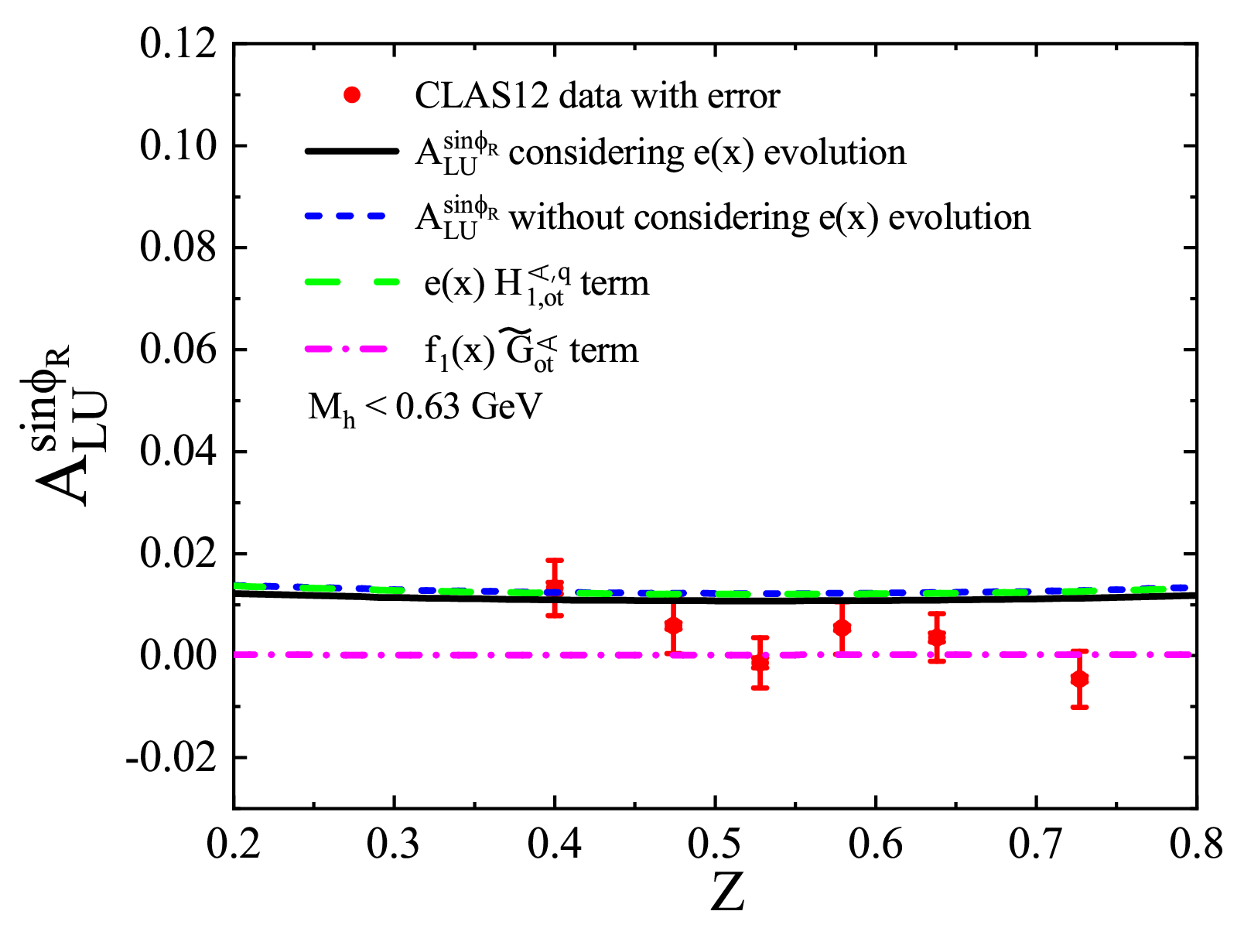}
  
\includegraphics[width=0.4\columnwidth]{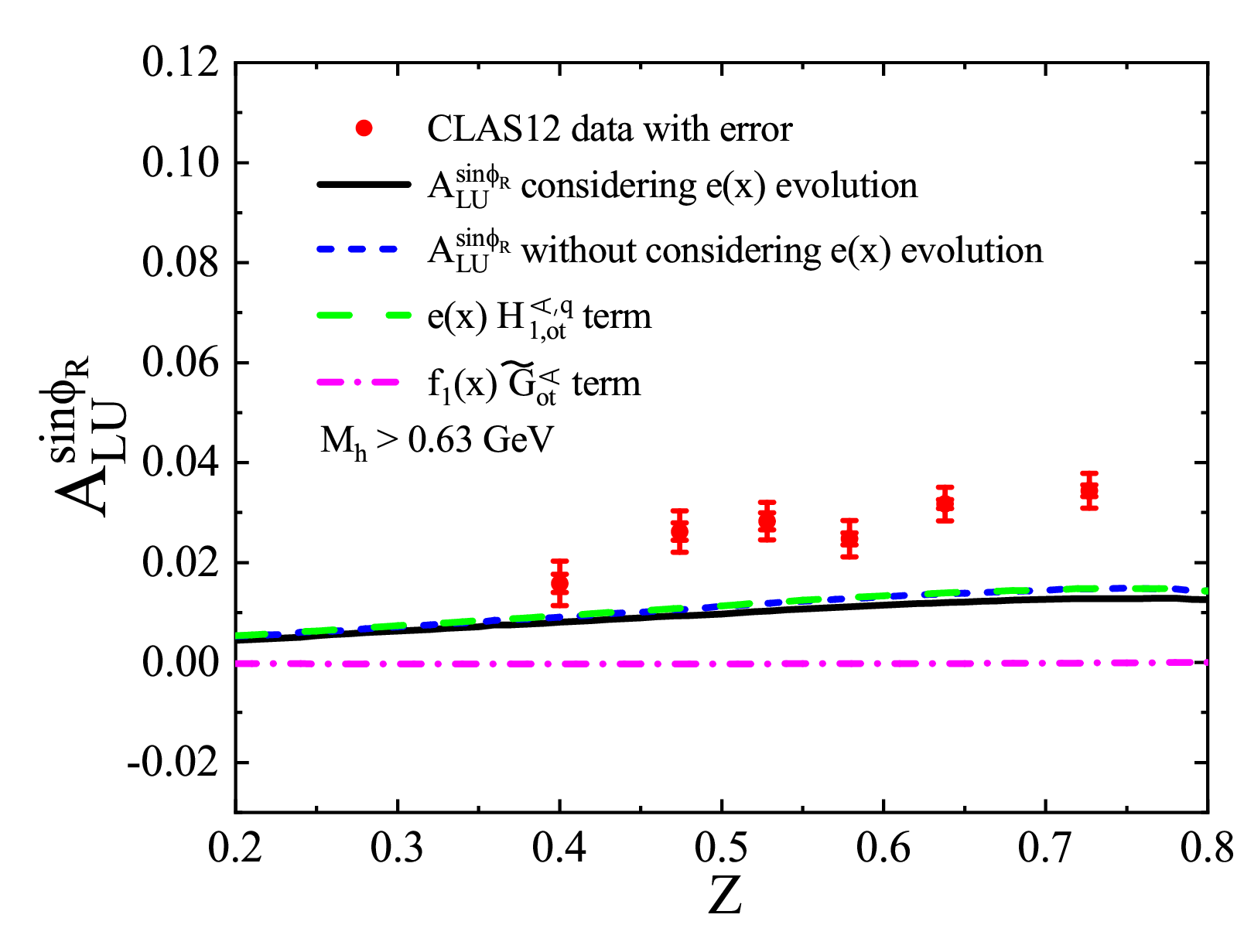}
  \includegraphics[width=0.4\columnwidth]{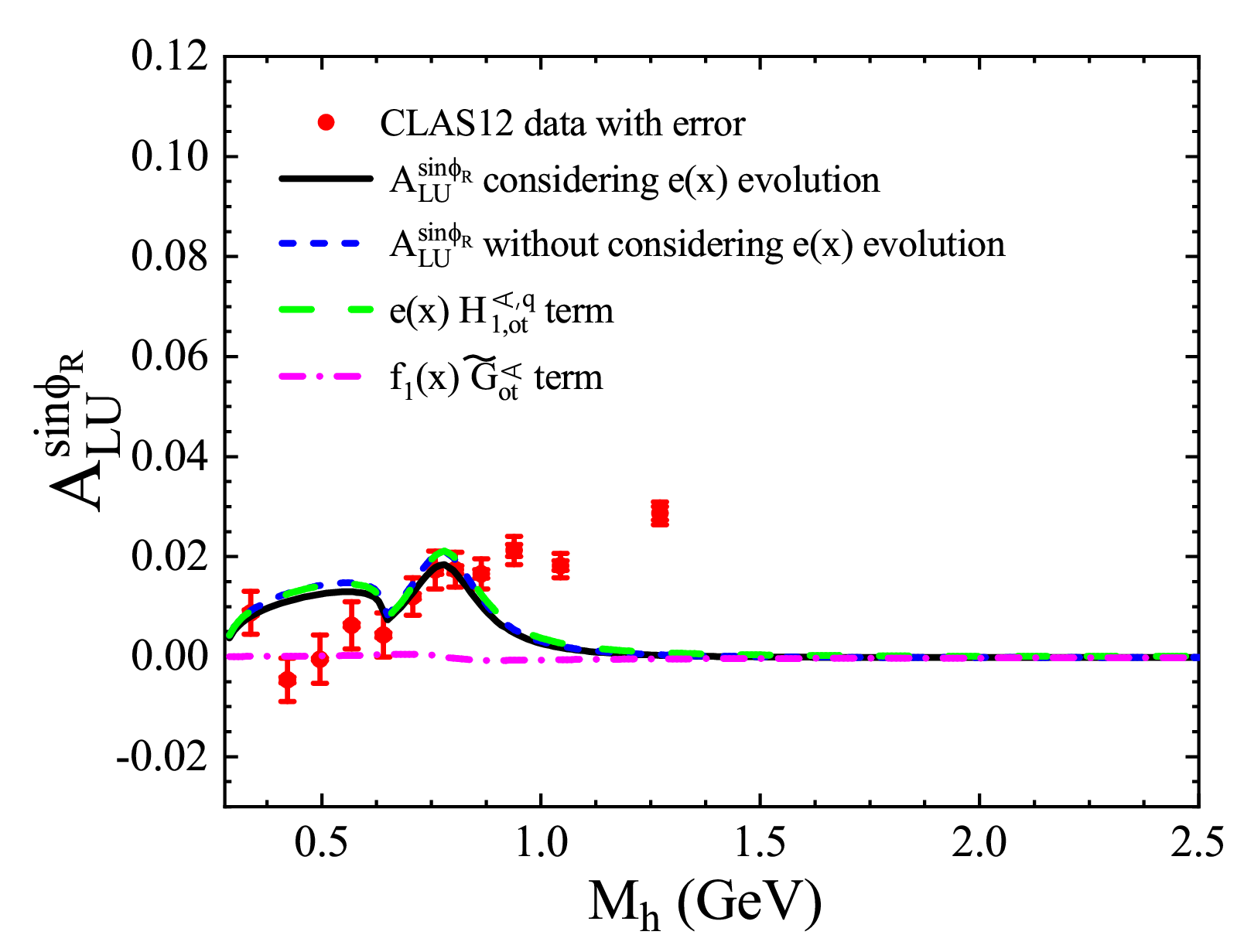}
  \caption{ Similar to Fig.~\ref{fig:asyCLAS} but at the kinematical region of CLAS12, the asymmetries are presented as functions of $x$ (upper left panel), $z$ with $M_h<0.63~\textrm{GeV}$(upper right panel), $z$ with $M_h>0.63~\textrm{GeV}$(lower left panel) and $M_h$ (lower right panel). The red full circles with error bars show the CLAS12 data~\cite{Hayward:2021psm} for comparison.}
  \label{fig:asyCLAS12}
\end{figure*}
 
To perform numerical calculations for the $\sin\phi_R$ azimuthal asymmetry of the dihadron production in SIDIS process at CLAS, we adopt the following kinematical cuts for CLAS~\cite{CLAS:2020igs}
\begin{align}
0.114<x<0.593,\quad 0.1<y&<0.85, \quad 0.530<z<0.948,\nonumber\\
0.279 ~\textrm{GeV} <M_h<1.734~\textrm{GeV}, \quad  1~\mathrm{GeV}^2&<Q^2<4.644~\mathrm{GeV}^2, \quad W>2~\mathrm{GeV}.
\label{eq:CLAScuts}
\end{align}
For CLAS12~\cite{Hayward:2021psm}, the following kinematic region is used
\begin{align}
0.06<x<0.77,\quad 0.1<y&<0.8, \quad 0.2<z<0.8, \nonumber\\
0.279~ \textrm{GeV} <M_h<2.5~\textrm{GeV}, \quad  1~\mathrm{GeV}^2&<Q^2<10.6~\mathrm{GeV}^2, \quad W> 2~\mathrm{GeV}.
\label{eq:CLAS12cuts}
\end{align}
Here, $W^2=(P+q)^2\approx\frac{1-x}{x}Q^2$ is the invariant mass of the virtual photon-nucleon system. In our calculation, we impose $W > 2 ~\textrm{GeV}$ to avoid the resonance region.
\begin{figure*} 
  \centering
  \includegraphics[width=0.32\columnwidth]{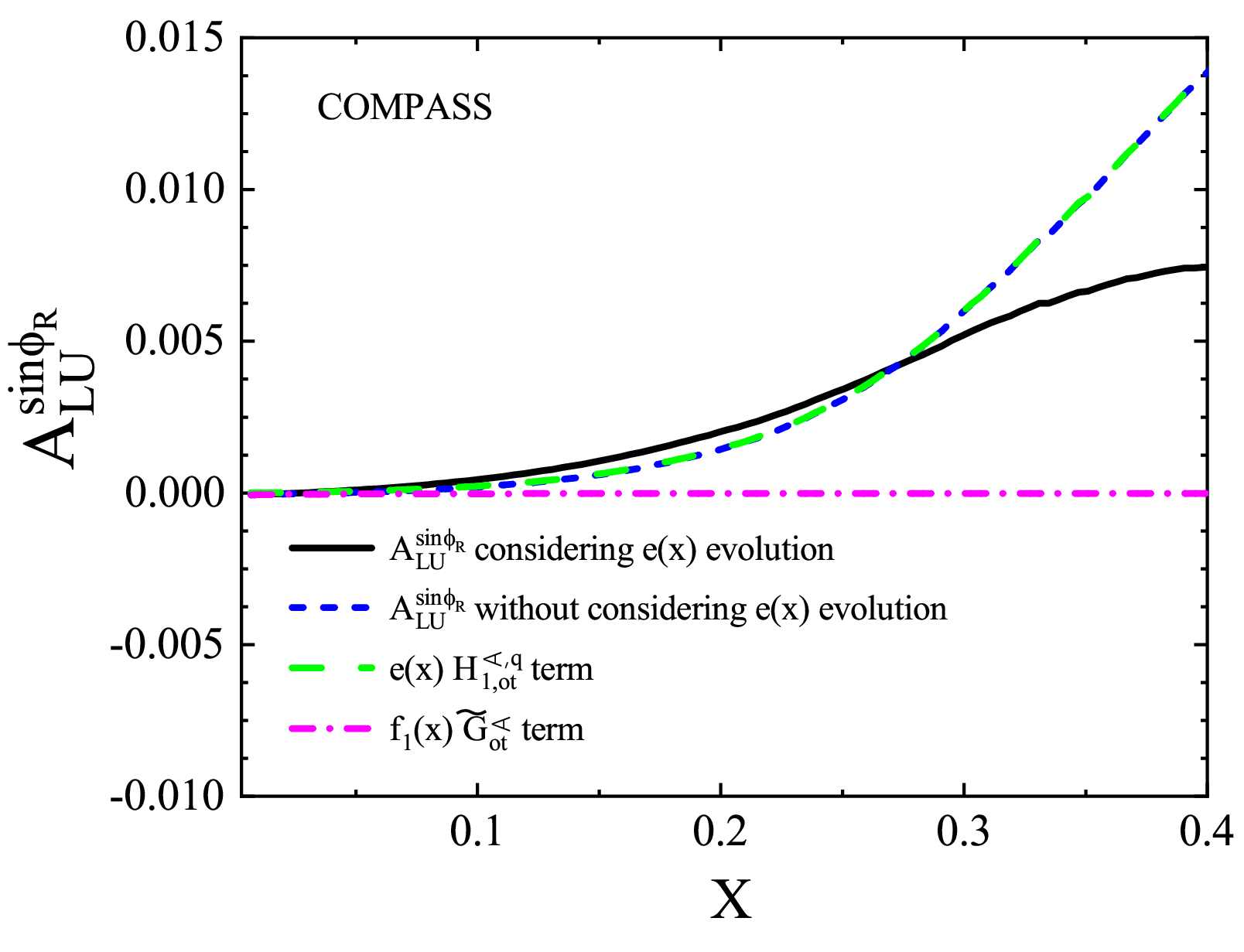}
  \includegraphics[width=0.32\columnwidth]{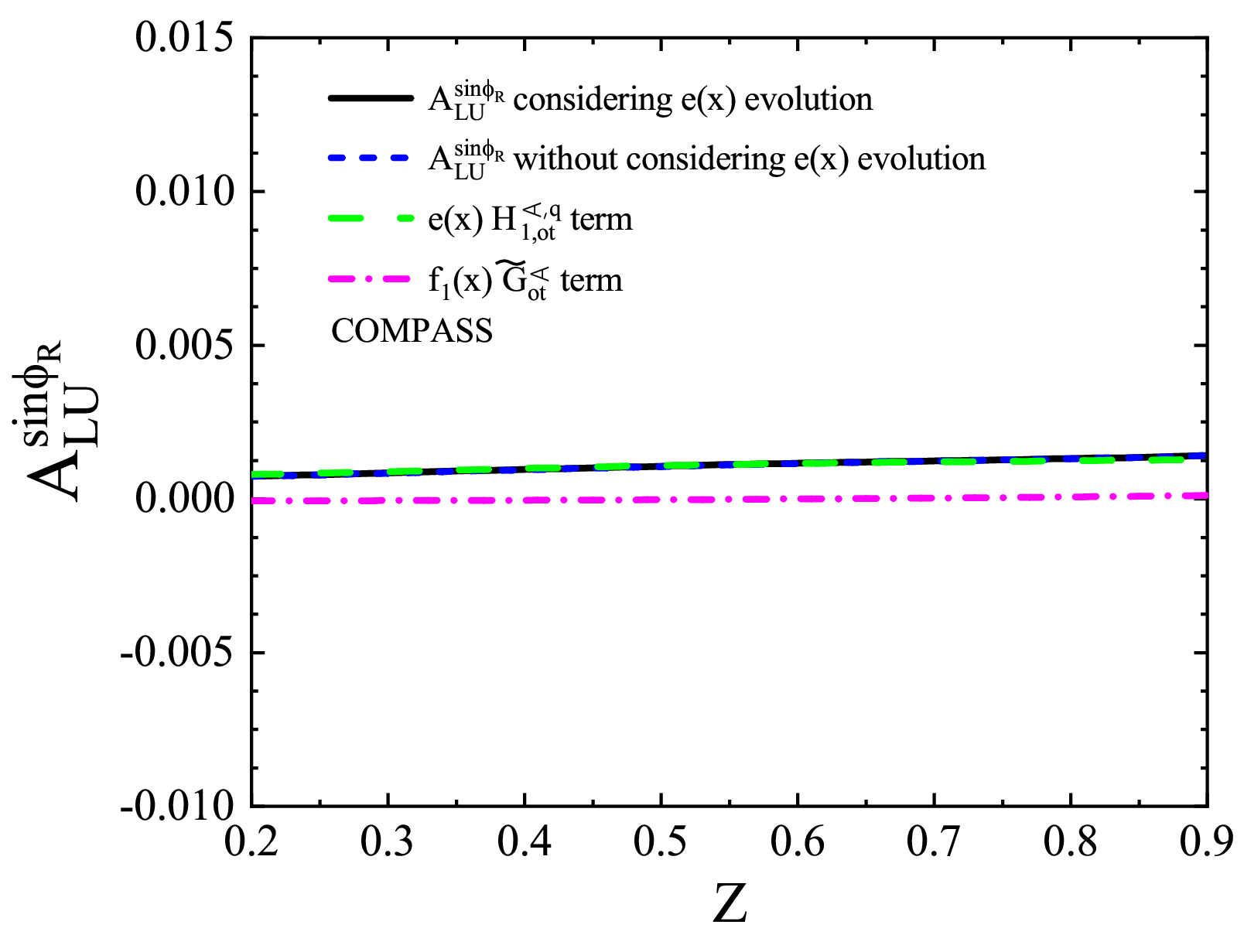}
  \includegraphics[width=0.32\columnwidth]{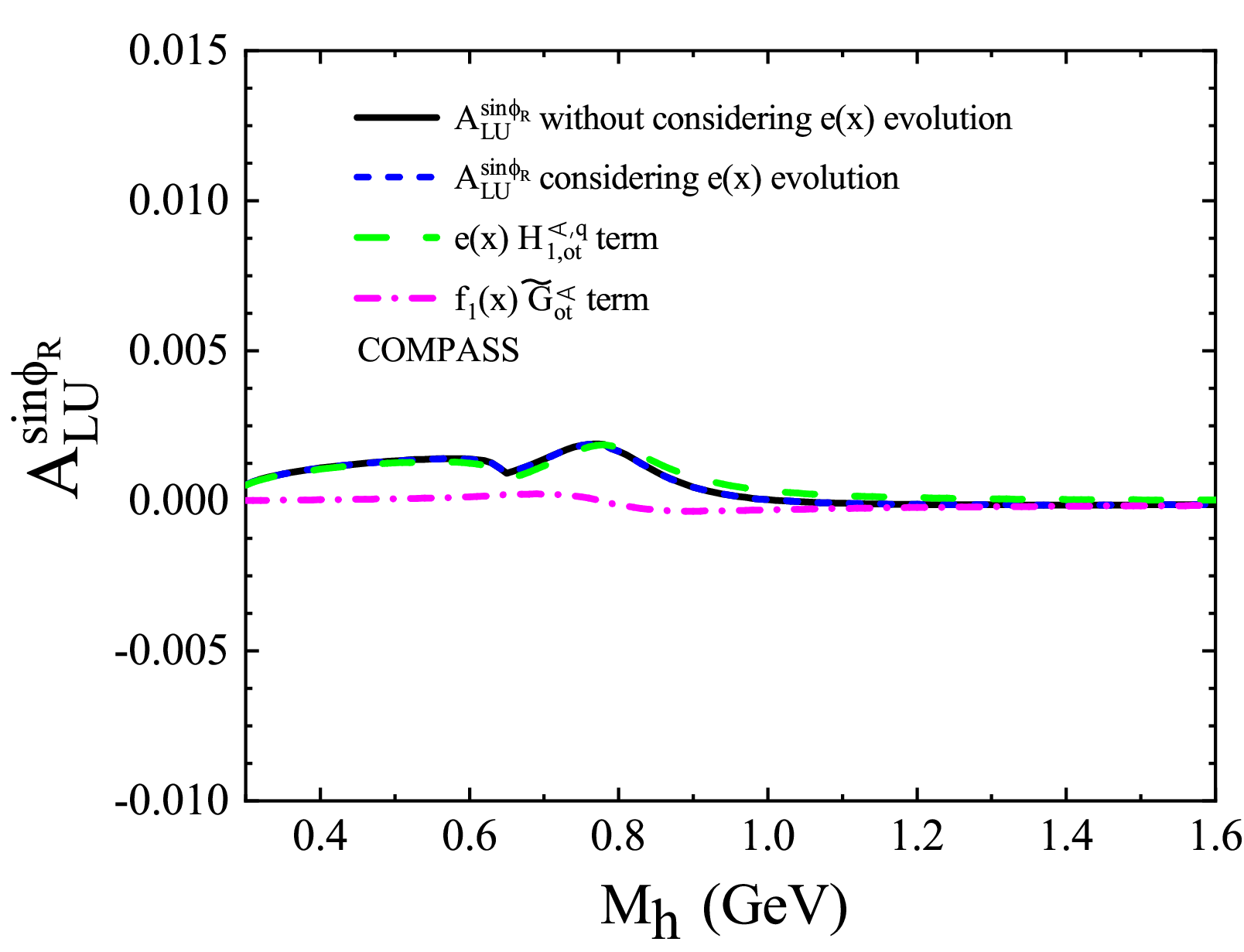}
  \includegraphics[width=0.32\columnwidth]{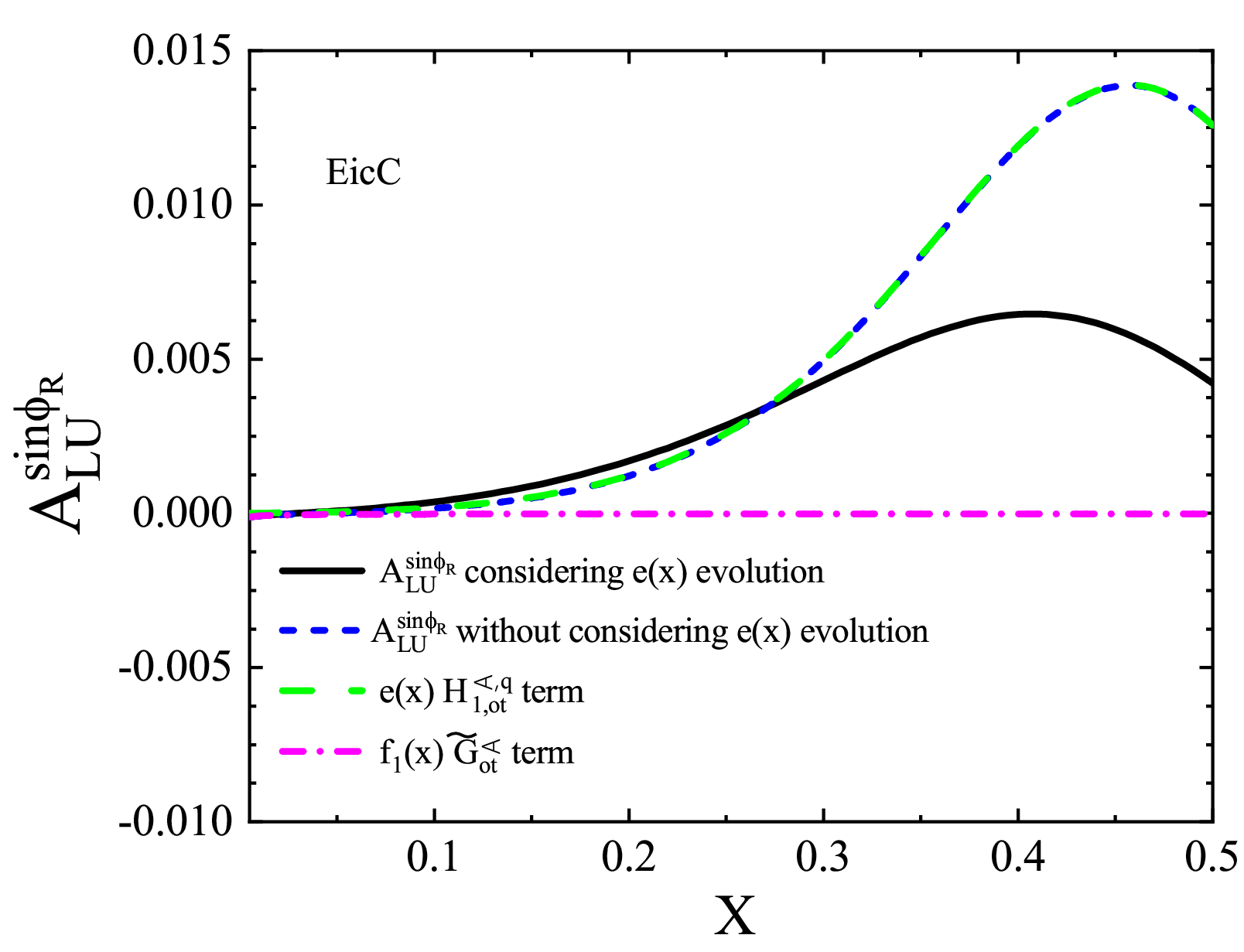}
  \includegraphics[width=0.32\columnwidth]{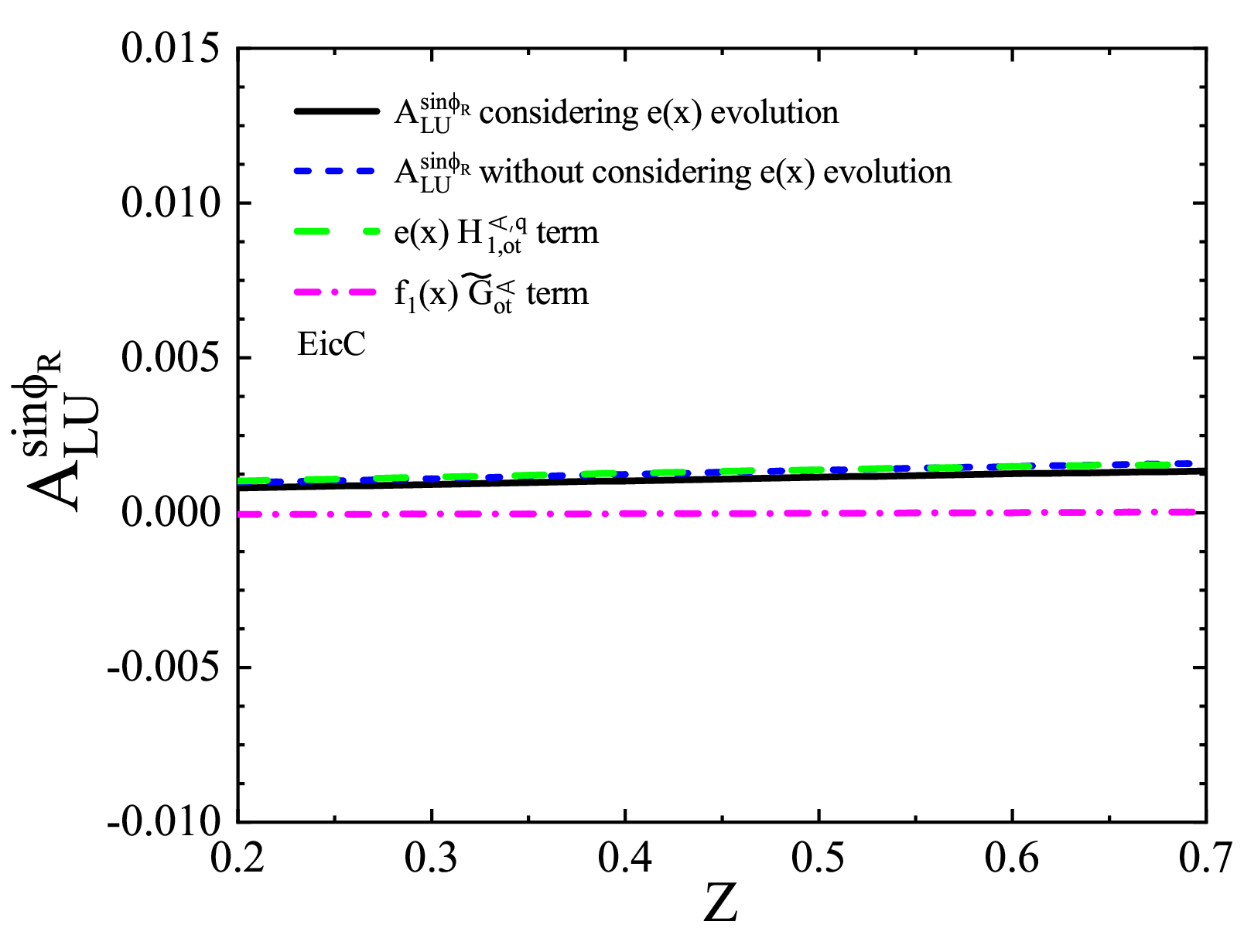}
  \includegraphics[width=0.32\columnwidth]{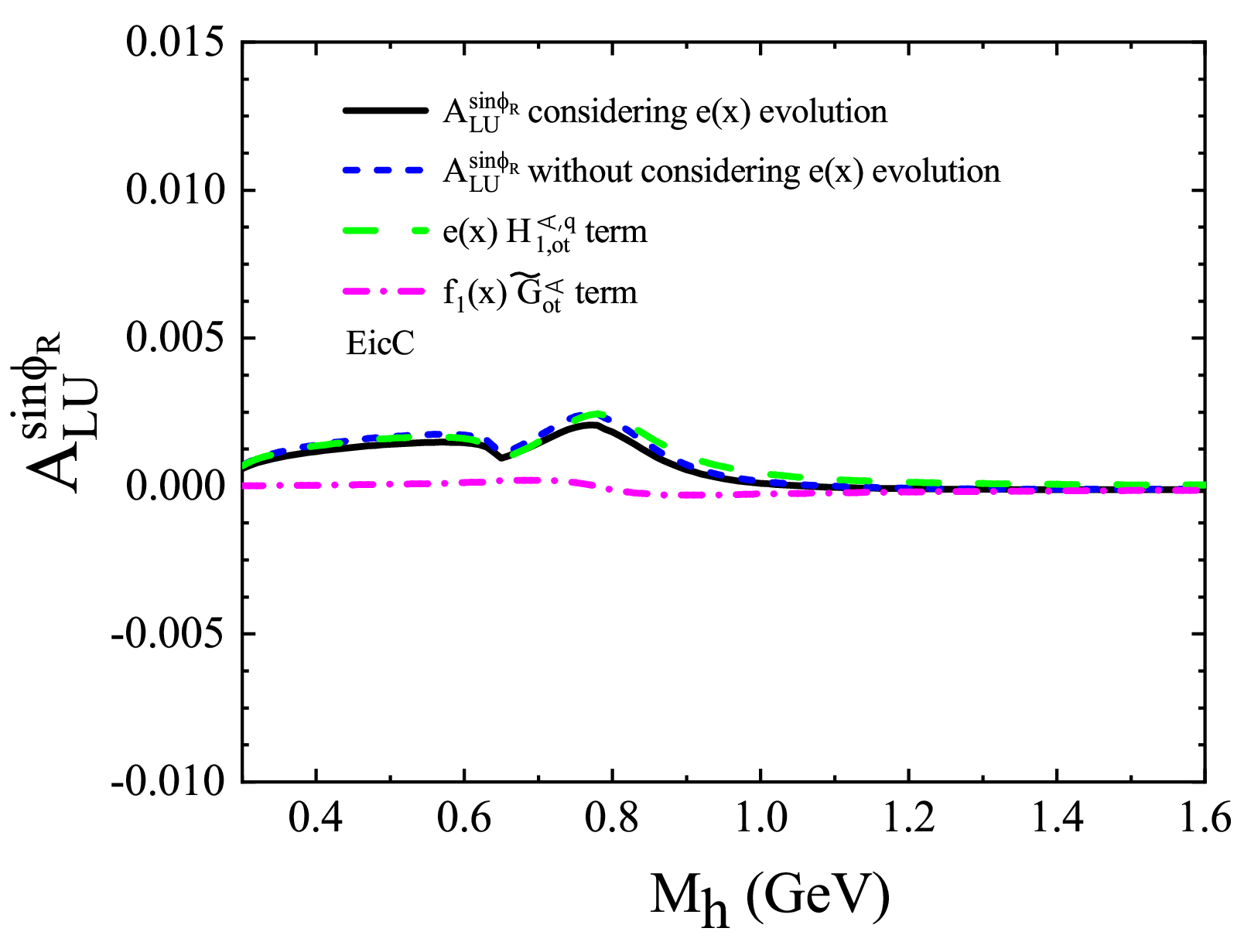}
 \includegraphics[width=0.32\columnwidth]{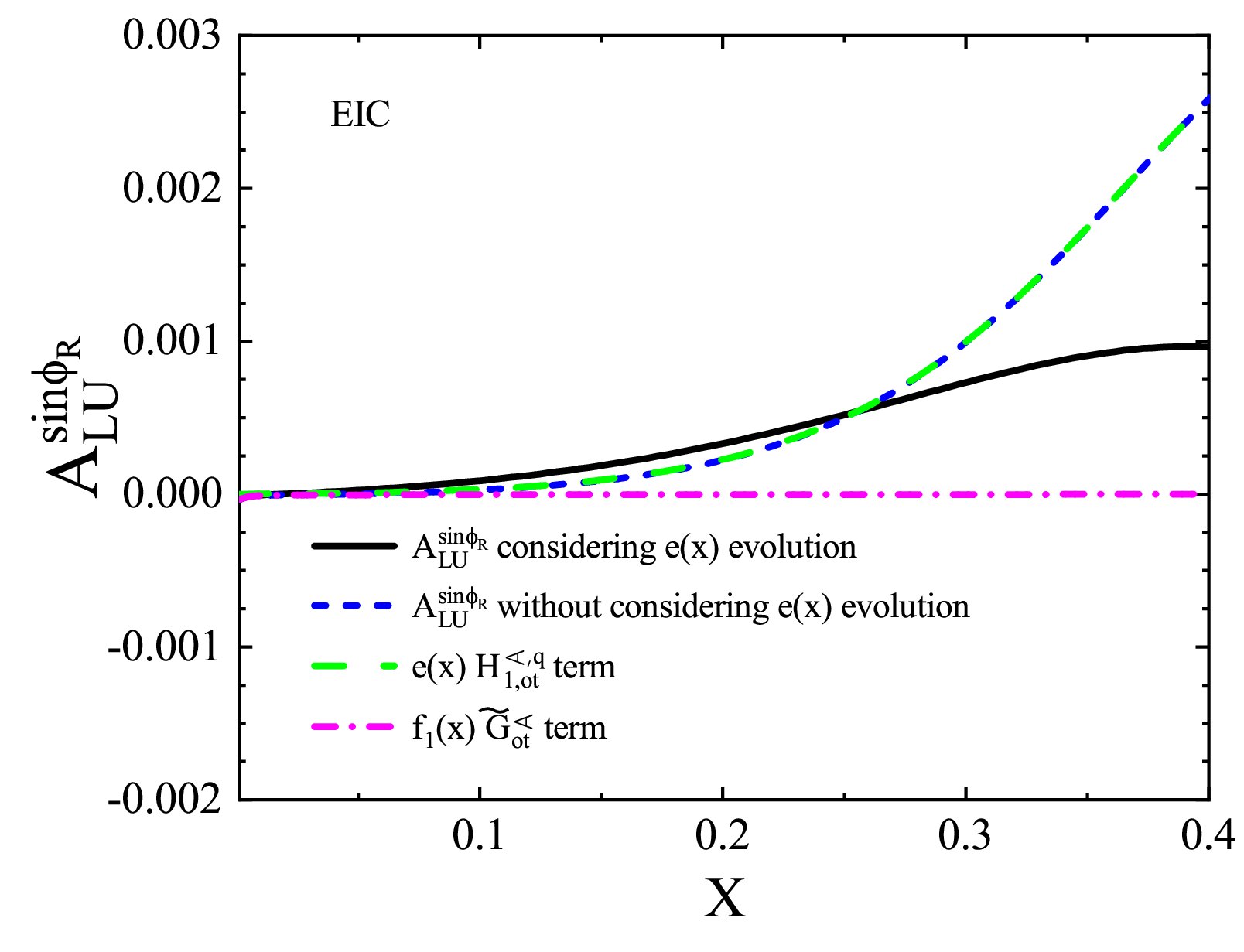}
  \includegraphics[width=0.32\columnwidth]{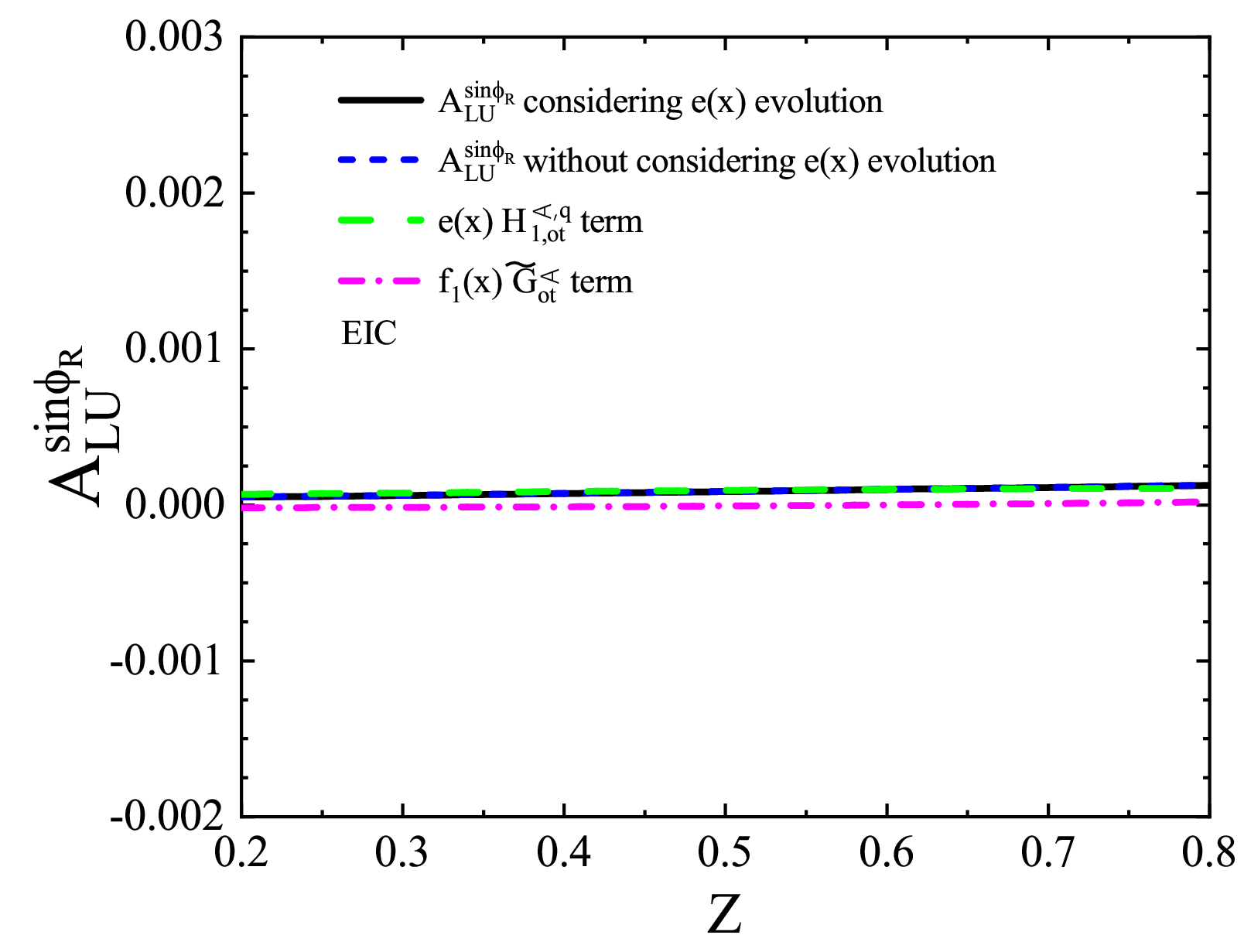}
  \includegraphics[width=0.32\columnwidth]{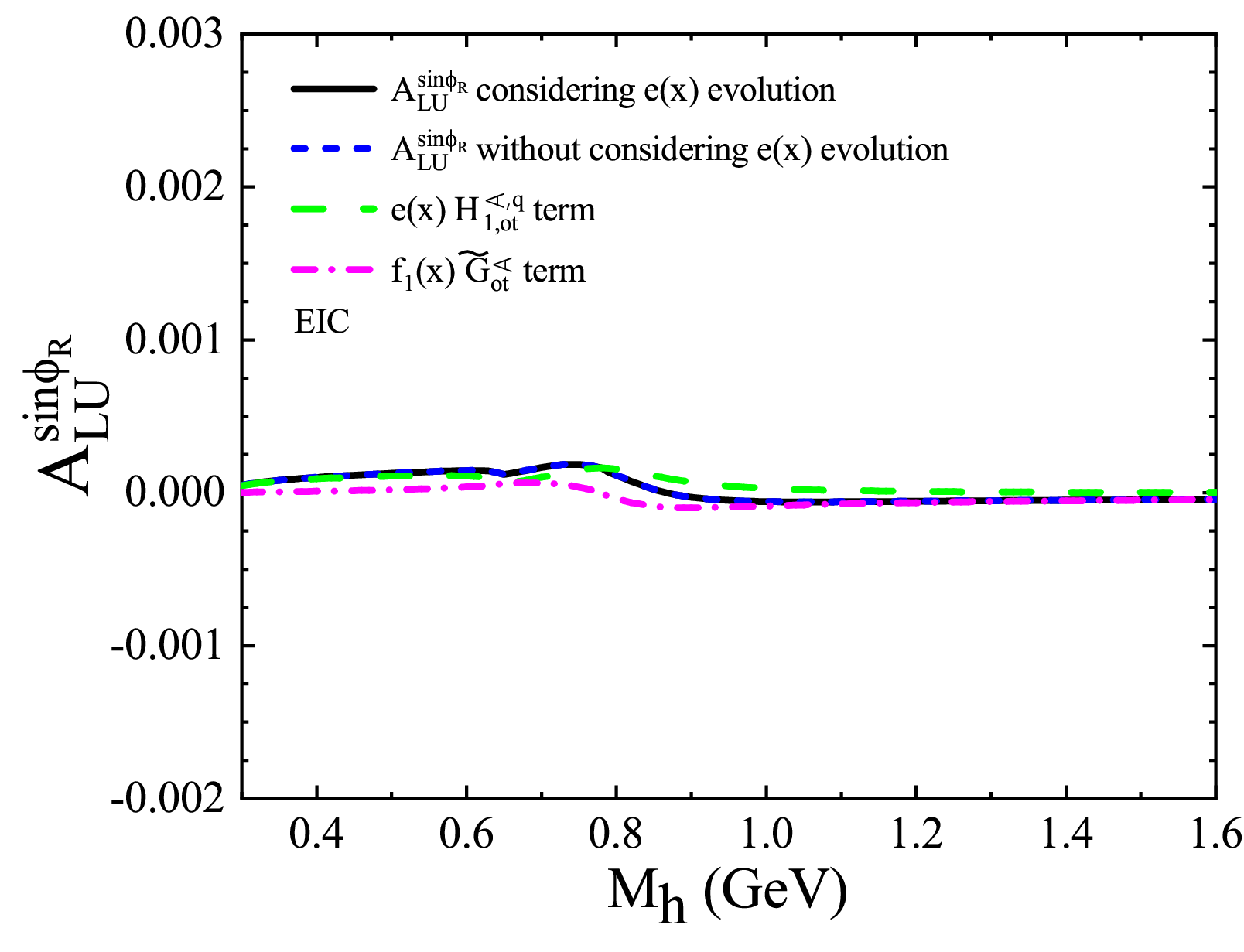}
  \caption{Similar to Fig.~\ref{fig:asyCLAS}, the $\sin\phi_R$ azimuthal asymmetries as functions of $x$ (left panel), $z$ (central panel) and $M_h$ (right panel) at the kinematics of the COMPASS, EicC, and EIC.}
  \label{fig:asyEIC,EIcC,COMPASS}
\end{figure*}

In Fig.~\ref{fig:asyCLAS}, we present the numerical results of the $\sin\phi_R$ azimuthal asymmetry in the SIDIS process of the dihadron production production with a longitudinally polarized electron beam scattered off an unpolarized proton the kinematical region of CLAS. The $x$-, $z$-, $M_h$- and $Q^2$-dependent asymmetries are shown in the upper-left, upper-right, lower-left, and lower-right panels, respectively.
Figure~\ref{fig:asyCLAS12} presents corresponding results for CLAS12 kinematics, with the $z$-dependence shown separately for two invariant mass regions ($M_h < 0.63$~GeV in upper-right and $M_h > 0.63$~GeV in lower-left panels), while the $x$- and $M_h$-dependencies appear in the upper-left and lower-right panels, respectively.

The dashed lines represent the contribution from the $e H_{1,ot}^{\sphericalangle}$ term, while the dash-dotted lines denote the contribution from the $f_1\,\widetilde{G}^{\sphericalangle}_{ot}$ term. The solid lines display the total contribution including the QCD evolution of $e(x)$, and the dashed lines show the total asymmetry without considering the QCD evolution, assuming that the DiFFs are scale independent. The experimental data from CLAS and CLAS12 are shown as circles with error bars.

The theoretical curves in Figs.~\ref{fig:asyCLAS} and~\ref{fig:asyCLAS12} reveal several important features.
For the $x$- and $Q^2$-dependent asymmetries, we find that including QCD evolution of $e(x)$ enhances the asymmetry in the small-$x$ and small-$Q^2$ regions, but reduces the asymmetry in the large-$x$ and large-$Q^2$ regions. This indicates that the scale dependence of $e(x)$ plays an important role in these observables. However, for the $z$- and $M_h$-dependent asymmetries, the effect of QCD evolution on $e(x)$ is negligible.
Moreover, the $eH_{1,ot}^{\sphericalangle}$ term is found to dominate the asymmetry, while the contribution from the $f_1\,\widetilde{G}^{\sphericalangle}_{ot}$ term is almost negligible and consistent with zero. Our theoretical results are in good agreement with the CLAS data, but deviate from the CLAS12 measurements, which could be attributed to the slightly higher $Q^2$ values in the CLAS12 kinematics.

In addition, we also make the predictions for the $\sin\phi_R$ asymmetries in beam single spin SIDIS process at the future COMPASS, EicC, and EIC. Such facilities would be ideal platforms for studying these observations.
For COMPASS~\cite{Sirtl:2017rhi}, based on the COMPASS convention, the depolarization factors are not included in the numerator and denominator. The energy of the lepton beam is $160~\textrm{GeV}$, which is obtained from Ref.~\cite{COMPASS:2023cgk}. For COMPASS, The following kinematical region is adopted
\begin{align}
&0.003<x<0.4,\quad 0.1<y<0.9, \quad 0.2<z<0.9,\nonumber\\
&0.3 ~\textrm{GeV} <M_h<1.6~\textrm{GeV}, \quad Q^2>1~\mathrm{GeV}^2,\quad W>5~\mathrm{GeV}.
\label{eq:COMPASScutS}
\end{align}
For EicC, the following kinematical region is chosen~\cite{Song:2024xar}
\begin{align}
&0.005<x<0.5,\quad 0.07<y<0.9, \quad 0.2<z<0.7,\nonumber\\
0.3 ~\textrm{GeV} <&M_h<1.6~\textrm{GeV}, \quad Q^2> 1~\mathrm{GeV}^2,\quad 
\sqrt{s}=16.7~\textrm{GeV}, \quad W>2~\mathrm{GeV}.
\label{eq:EIcCcuts}
\end{align}
And for EIC
~\cite{Matevosyan:2015gwa,Accardi:2012qut}:
\begin{align}
&0.001<x<0.4,\quad 0.01<y<0.95, \quad 0.2<z<0.8,\nonumber\\
0.3 ~\textrm{GeV} <&M_h<1.6~\textrm{GeV}, \quad Q^2> 1~\mathrm{GeV}^2,\quad 
\sqrt{s}=100~\textrm{GeV},\quad W>
5~\mathrm{GeV}.
\label{eq:EICcuts}
\end{align}

The $x$-, $z$- and $M_h$-dependent asymmetries are plotted in the left, central, and right panels of Fig.~\ref{fig:asyEIC,EIcC,COMPASS}, respectively. The overall tendencies of the asymmetries at the COMPASS, EicC, and EIC are similar to those at CLAS and CLAS12, though with reduced magnitudes due to twist-3 observable suppression at higher $Q^2$. Despite this suppression, the asymmetries remain measurable across all three facilities.

It also shows that including the QCD evolution of $e(x)$ would change the $x$-dependent asymmetries, which indicates that the scale dependence of $e(x)$ plays an essential role in these observables. In addition, the asymmetry contributed by the coupling of $f_1(x)$ and $\widetilde{G}^{\sphericalangle}$ is consistent with zero in all kinematical regions, which confirms the validity of the Wandzura-Wilczek approximation in the study of BSA.

\section{CONCLUSION}
\label{Sec.conclusion}

In this work, we have investigated the BSA $A^{\sin\phi_R}_{LU}$ in the dihadron production through SIDIS process with a longitudinally polarized electron beam scattered off the unpolarized proton target. 
The analysis incorporates two contributions. One is the coupling between twist-3 PDF $e(x)$ and the twist-2 DiFF $H_{1,ot}^{\sphericalangle}$; the other is the coupling between the unpolarized PDF $f_1(x)$ and the twist-3 DiFF $\widetilde{G}^{\sphericalangle}_{ot}$.
Employing the spectator model results for the relevant DiFFs and twist-3 PDF, we estimated the BSA $A^{\sin \phi_R}_{LU}$ and compared it with the CLAS and CLAS12 measurements.
Our numerical calculations demonstrate good agreement with the CLAS measurements for the $z$- and $M_h$- dependent asymmetries, though discrepancies emerge in the $x$- and $Q^2$- dependent asymmetries. 
Similar qualitative behavior persists in CLAS12 kinematics.

The impact of the DGLAP evolution effect of $e(x)$  on the BSA was systematically studied, revealing reduced asymmetries in the higher $x$ and $Q^2$ regimes, while the $z$- and $M_h$- dependent asymmetries remain largely unaffected by the evolution effect of $e(x)$.
In addition, we predicted the BSA $A^{\sin \phi_R}_{LU}$ in dihadron production at the kinematical regions of COMPASS, EIC, and EicC, showing measurable asymmetries with suppressed magnitudes due to higher-twist nature of the observable at increased $Q^2$ scales.

Our results indicate that the $e(x)\,H_{1,ot}^{\sphericalangle}$ term dominates across most kinematical regions, while the contribution from the $f_1(x)\,\widetilde{G}^{\sphericalangle}_{ot}$ term is consistent with zero -- a result consistent with the Wandzura-Wilczek approximation in the study of BSA. 
These results underscore the importance of the QCD evolution effect in twist-3 PDF analyses and the Wandzura-Wilczek approximation in future phenomenological studies of the BSA $A^{\sin \phi_R}_{LU}$. Such refinements will enhance precision in extracting $e(x)$ and advance our understanding of nucleon structure through polarized SIDIS measurements.

\section{ACKNOWLEDGMENTS}

X. Wang is supported by the Natural Science Foundation of Henan Province under Grant Nos. 242300421377 and 232300421140. Hui Li is supported by the Fundamental Research Program of Shanxi Province (No. 202203021222224) and the Natural Science Foundation of Shanxi Normal University (No. JCYJ2023021).


\begin{thebibliography}{99}

\bibitem{Metz:2016swz}
A.~Metz and A.~Vossen,
Prog. Part. Nucl. Phys. \textbf{91}, 136-202 (2016).

\bibitem{Wen:2024cfu}
X.~K.~Wen, B.~Yan, Z.~Yu and C.~P.~Yuan, (2024),
arXiv:2408.07255 [hep-ph].

\bibitem{Boussarie:2023izj}
R.~Boussarie, M.~Burkardt and M.~Constantinou, \textit{et al.}, (2023),
arXiv:2304.03302 [hep-ph].

\bibitem{Sharma:2023wha}
S.~Sharma, N.~Kumar and H.~Dahiya,
Nucl. Phys. B \textbf{992}, 116247 (2023).

\bibitem{Jaffe:1991ra}
R.~L.~Jaffe and X.~D.~Ji,
Nucl. Phys. \textbf B \textbf{375}, 527-560  (1992).

\bibitem{Wakamatsu:2000fd}
M.~Wakamatsu,
Phys. Lett. B \textbf{509}, 59-68 (2001).

\bibitem{Wakamatsu:2003uu}
M.~Wakamatsu and Y.~Ohnishi,
Phys. Rev. D \textbf{67}, 114011  (2003).

\bibitem{Schweitzer:2003uy}
P.~Schweitzer,
Phys. Rev. D \textbf{67}, 114010 (2003).


\bibitem{Mukherjee:2009uy}
A.~Mukherjee,
Phys. Lett. B \textbf{687}, 180-183  (2010).


\bibitem{Avakian:2010br}
H.~Avakian, A.~V.~Efremov, P.~Schweitzer and F.~Yuan,
Phys. Rev. D \textbf{81}, 074035 (2010).

\bibitem{Lorce:2014hxa}
C.~Lorc\'e, B.~Pasquini and P.~Schweitzer,
JHEP \textbf{01}, 103 (2015).

\bibitem{Pasquini:2018oyz}
B.~Pasquini and S.~Rodini,
Phys. Lett. B \textbf{788}, 414-424 (2019).

\bibitem{Bastami:2020rxn}
S.~Bastami, A.~V.~Efremov and P.~Schweitzer, \textit{et al.}
Phys. Rev. D \textbf{103}, 014024 (2021).


\bibitem{Ji:2020baz}
X.~Ji,
Nucl. Phys. \textbf{B}, 115181 (2020).

\bibitem{Ji:1994av}
X.~D.~Ji,
Phys. Rev. Lett. \textbf{74}, 1071-1074 (1995).

\bibitem{Ji:2021mtz}
X.~Ji,
Front. Phys. (Beijing) \textbf{16}, 64601 (2021).

\bibitem{Lorce:2017xzd}
C.~Lorc\'e,
Eur. Phys. J. C \textbf{78}, 120 (2018).

\bibitem{Lorce:2021xku}
C.~Lorc\'e, A.~Metz, B.~Pasquini and S.~Rodini,
JHEP \textbf{11}, 121 (2021).

\bibitem{AbdulKhalek:2021gbh}
R.~Abdul Khalek, A.~Accardi and J.~Adam, \textit{et al.}
Nucl. Phys. \textbf A \textbf{1026}, 122447 (2022).

\bibitem{Burkardt:2008ps}
M.~Burkardt,
Phys. Rev. D \textbf{88}, 114502 (2013).

\bibitem{Gamberg:2003pz}
L.~P.~Gamberg, D.~S.~Hwang and K.~A.~Oganessyan,
Phys. Lett. B \textbf{584}, 276-284 (2004).

\bibitem{Jakob:1997wg}
R.~Jakob, P.~J.~Mulders and J.~Rodrigues,
Nucl. Phys. \textbf A \textbf{626}, 937-965 (1997).

\bibitem{HERMES:2006pof}
A.~Airapetian \textit{et al.} [HERMES],
Phys. Lett. B \textbf{648}, 164-170 (2007).

\bibitem{CLAS:2014dmz}
W.~Gohn \textit{et al.} [CLAS],
Phys. Rev. D \textbf{89}, 072011 (2014).

\bibitem{Moretti:2019lkw}
A.~Moretti [COMPASS],
PoS \textbf{SPIN2018}, 052 (2019).

\bibitem{Efremov:2002ut}
A.~V.~Efremov, K.~Goeke and P.~Schweitzer,
Phys. Rev. D \textbf{67}, 114014 (2003). 

\bibitem{CLAS:2003qum}
H.~Avakian \textit{et al.} [CLAS],
Phys. Rev. D \textbf{69}, 112004 (2004).

\bibitem{Courtoy:2022kca}
A.~Courtoy, A.~S.~Miramontes and H.~Avakian, \textit{et al.} 
Phys. Rev. D \textbf{106}, 014027 (2022).

\bibitem{Konishi:1978yx}
K.~Konishi, A.~Ukawa and G.~Veneziano,
Phys. Lett. B \textbf{78}, 243-248 (1978).

\bibitem{Vendramin:1980wz}
I.~Vendramin,
Nuovo Cim. A \textbf{62}, 21 (1981).

\bibitem{Vendramin:1981te}
I.~Vendramin,
Nuovo Cim. A \textbf{66}, 339 (1981).

\bibitem{Ceccopieri:2007ip}
F.~A.~Ceccopieri, M.~Radici and A.~Bacchetta,
Phys. Lett. B \textbf{650}, 81-89 (2007).

\bibitem{Collins:1994ax}
J.~C.~Collins and G.~A.~Ladinsky, (1994),
arXiv:hep-ph/9411444 [hep-ph].

\bibitem{Bianconi:1999cd}
A.~Bianconi, S.~Boffi, R.~Jakob and M.~Radici,
Phys. Rev. D \textbf{62}, 034008 (2000).

\bibitem{Radici:2001na}
M.~Radici, R.~Jakob and A.~Bianconi,
Phys. Rev. D \textbf{65}, 074031 (2002).

\bibitem{Bacchetta:2003vn}
A.~Bacchetta and M.~Radici,
Phys. Rev. D \textbf{69}, 074026 (2004).



\bibitem{Bacchetta:2002ux}
A.~Bacchetta and M.~Radici,
Phys. Rev. D \textbf{67}, 094002 (2003). 

\bibitem{Courtoy:2012ry}
A.~Courtoy, A.~Bacchetta, M.~Radici and A.~Bianconi,
Phys. Rev. D \textbf{85}, 114023 (2012).

\bibitem{Bianconi:1999uc}
A.~Bianconi, S.~Boffi, R.~Jakob and M.~Radici,
Phys. Rev. D \textbf{62}, 034009 (2000).

\bibitem{Bacchetta:2006un}
A.~Bacchetta and M.~Radici,
Phys. Rev. D \textbf{74}, 114007 (2006).


\bibitem{Bacchetta:2008wb}
A.~Bacchetta, F.~A.~Ceccopieri, A.~Mukherjee and M.~Radici,
Phys. Rev. D \textbf{79}, 034029 (2009).

\bibitem{Matevosyan:2013aka}
H.~H.~Matevosyan, A.~W.~Thomas and W.~Bentz,
Phys. Rev. D \textbf{88}, 094022 (2013).

\bibitem{Matevosyan:2013eia}
H.~H.~Matevosyan, A.~Kotzinian and A.~W.~Thomas,
Phys. Lett. B \textbf{731}, 208-216 (2014).


\bibitem{Matevosyan:2017alv}
H.~H.~Matevosyan, A.~Kotzinian and A.~W.~Thomas,
Phys. Rev. D \textbf{96}, 074010 (2017). 

\bibitem{Matevosyan:2017uls}
H.~H.~Matevosyan, A.~Kotzinian and A.~W.~Thomas,
Phys. Rev. D \textbf{97}, 014019 (2018).

\bibitem{Diehl:2023nmm}
S.~Diehl,
Prog. Part. Nucl. Phys. \textbf{133}, 104069 (2023).

\bibitem{CLAS:2020igs}
M.~Mirazita \textit{et al.} [CLAS],
Phys. Rev. Lett. \textbf{126}, 062002 (2021).

\bibitem{Hayward:2021psm}
T.~B.~Hayward, C.~Dilks, A.~Vossen and H.~Avakian, \textit{et al.}
Phys. Rev. Lett. \textbf{126}, 152501  (2021).

\bibitem{Yang:2019aan}
W.~Yang, X.~Wang, Y.~Yang and Z.~Lu,
Phys. Rev. D \textbf{99}, 054003 (2019).

\bibitem{Mao:2013waa}
W.~Mao and Z.~Lu,
Eur. Phys. J. C \textbf{73}, 2557 (2013).

\bibitem{Boer:2003cm}
D.~Boer, P.~J.~Mulders and F.~Pijlman,
Nucl. Phys. \textbf B \textbf{667}, 201-241 (2003).

\bibitem{Lu:2015wja}
Z.~Lu and I.~Schmidt,
Phys. Lett. B \textbf{747}, 357-364 (2015).

\bibitem{Lai:2010vv}
H.~L.~Lai, M.~Guzzi and J.~Huston, \textit{et al.} 
Phys. Rev. D \textbf{82}, 074024 (2010).

\bibitem{Mao:2012dk}
W.~Mao and Z.~Lu,
Phys. Rev. D \textbf{87}, 014012 (2013).

\bibitem{CLAS:2021opg}
S.~Diehl \textit{et al.} [CLAS],
Phys. Rev. Lett. \textbf{128}, 062005 (2022).

\bibitem{Bacchetta:2003rz}
A.~Bacchetta, A.~Schaefer and J.~J.~Yang,
Phys. Lett. B \textbf{578}, 109-118 (2004).

\bibitem{Bacchetta:2008af}
A.~Bacchetta, F.~Conti and M.~Radici,
Phys. Rev. D \textbf{78}, 074010 (2008).

\bibitem{Salam:2008qg}
G.~P.~Salam and J.~Rojo,
Comput. Phys. Commun. \textbf{180}, 120-156 (2009).


\bibitem{Gamberg:2003eg}
L.~P.~Gamberg, G.~R.~Goldstein and K.~A.~Oganessyan,
Phys. Rev. D \textbf{68}, 051501 (2003).

\bibitem{Bacchetta:2002tk}
A.~Bacchetta, R.~Kundu, A.~Metz and P.~J.~Mulders,
Phys. Rev. D \textbf{65}, 094021 (2002).

\bibitem{Yang:2017cwi}
Y.~Yang, Z.~Lu and I.~Schmidt,
Phys. Rev. D \textbf{96}, 034010 (2017).

\bibitem{Belitsky:2002sm}
A.~V.~Belitsky, X.~Ji and F.~Yuan,
Nucl. Phys. \textbf B \textbf{656}, 165-198 (2003).

\bibitem{Yang:2016mxl}
Y.~Yang, Z.~Lu and I.~Schmidt,
Phys. Lett. B \textbf{761}, 333-339 (2016).

\bibitem{Amrath:2005gv}
D.~Amrath, A.~Bacchetta and A.~Metz,
Phys. Rev. D \textbf{71}, 114018 (2005). 

\bibitem{Singh:2005sbx}
B.~Singh,
Nucl. Data Sheets \textbf{106}, 601-618 (2005).

\bibitem{Courtoy:2014ixa}
A.~Courtoy, (2014),
arXiv:1405.7659 [hep-ph].

\bibitem{Sirtl:2017rhi}
S.~Sirtl, (2017),
arXiv:1702.07317 [hep-ex].

\bibitem{COMPASS:2023cgk}
G.~D.~Alexeev \textit{et al.} [COMPASS],
Phys. Lett. B \textbf{845}, 138155 (2023).

\bibitem{Song:2024xar}
J.~Song, Y.~Li, S.~C.~Xue, H.~Li and X.~Wang,
Universe \textbf{10}, 280 (2024).

\bibitem{Matevosyan:2015gwa}
H.~H.~Matevosyan, A.~Kotzinian and E.~C.~Aschenauer, \textit{et al.}
Phys. Rev. D \textbf{92}, 054028 (2015).

\bibitem{Accardi:2012qut}
A.~Accardi, J.~L.~Albacete, M.~Anselmino,  \textit{et al.}
Eur. Phys. J. A \textbf{52}, 268 (2016).

\end{thebibliography}
\end{document}